\documentclass[twocolumn,notitlepage,prb,color,superscriptaddress,psfig,showpacs,amsmath,amssymb,nobibnotes,longbibliography]{revtex4-2}

\usepackage{xspace}
\usepackage{color}
\usepackage{amsmath}
\usepackage{amssymb,stmaryrd}   
\usepackage{mathrsfs}
\usepackage{graphicx,subfigure}     
\usepackage{comment}
\usepackage{indentfirst}

\definecolor{viol}{rgb}{0.7, 0.4, 1}

\usepackage{placeins}
\usepackage{soul}

\definecolor{mygray}{cmyk}{0, 0, 0, 0.3}

\usepackage{float}
\usepackage[colorlinks=true,citecolor=blue]{hyperref}

\definecolor{darkgreen}{rgb}{0, 0.4, 0}

\newcommand{\CCPS}{CuCrP$_{2}$S$_{6}$\xspace}
\newcommand{\MPS}{Mn$_{2}$P$_{2}$S$_{6}$\xspace}

\newcommand{\NPS}{Ni$_{2}$P$_{2}$S$_{6}$\xspace}
\newcommand{\FPS}{Fe$_{2}$P$_{2}$S$_{6}$\xspace}

\newcommand{\CGT}{Cr$_{\text{2}}$Ge$_{\text{2}}$Te$_{\text{6}}$\xspace}

\newcommand{\HIIa}{$\textbf{H} \parallel {\bf a}$\xspace}
\newcommand{\HIIb}{$\textbf{H} \parallel {\bf b}$\xspace}
\newcommand{\HIIc}{$\textbf{H} \parallel {\bf c^*}$\xspace}

\newcommand{\degree}{^\circ}

\begin{document}

\title{Magnetic-field tuning of the spin dynamics in the quasi-2D van der Waals antiferromagnet \CCPS}

\author{Joyal John Abraham}
\thanks{These authors contributed equally to this work.}
\affiliation{Leibniz Institute for Solid State and Materials Research, Helmholtzstr. 20, D-01069 Dresden, Germany}
\affiliation{Institute for Solid State and Materials Physics, TU Dresden, D-01062 Dresden, Germany}

\author{Yaqian Guo}
\thanks{These authors contributed equally to this work.}
\affiliation{Leibniz Institute for Solid State and Materials Research, Helmholtzstr. 20, D-01069 Dresden, Germany}
\affiliation{Institute for Solid State and Materials Physics, TU Dresden, D-01062 Dresden, Germany}

\author{Yuliia Shemerliuk}
\affiliation{Leibniz Institute for Solid State and Materials Research, Helmholtzstr. 20, D-01069 Dresden, Germany}
\affiliation{Institute for Solid State and Materials Physics, TU Dresden, D-01062 Dresden, Germany}

\author{Sebastian Selter}
\affiliation{Leibniz Institute for Solid State and Materials Research, Helmholtzstr. 20, D-01069 Dresden, Germany}

\author{Saicharan Aswartham}
\affiliation{Leibniz Institute for Solid State and Materials Research, Helmholtzstr. 20, D-01069 Dresden, Germany}

\author{Kranthi Kumar Bestha}
\affiliation{Leibniz Institute for Solid State and Materials Research, Helmholtzstr. 20, D-01069 Dresden, Germany}

\author{Laura T. Corredor}
\thanks{Present address: Faculty of Physics, Technical University of Dortmund,
Otto-Hahn-Str. 4, D-44227 Dortmund, Germany.}
\affiliation{Leibniz Institute for Solid State and Materials Research, Helmholtzstr. 20, D-01069 Dresden, Germany}

\author{Anja U. B. Wolter}
\affiliation{Leibniz Institute for Solid State and Materials Research, Helmholtzstr. 20, D-01069 Dresden, Germany}

\author{Olga Kataeva}
\affiliation{Leibniz Institute for Solid State and Materials Research, Helmholtzstr. 20, D-01069 Dresden, Germany}

\author{Luka Rogi\'c}
\affiliation{Department of Physics, Faculty of Science, University of Zagreb, Bijeni\v{c}ka 32, HR-10000 Zagreb, Croatia}

\author{Noah Somun}
\affiliation{Department of Physics, Faculty of Science, University of Zagreb, Bijeni\v{c}ka 32, HR-10000 Zagreb, Croatia}

\author{Damjan Pelc}
\affiliation{Department of Physics, Faculty of Science, University of Zagreb, Bijeni\v{c}ka 32, HR-10000 Zagreb, Croatia}

\author{Oleg Janson}
\affiliation{Leibniz Institute for Solid State and Materials Research, Helmholtzstr. 20, D-01069 Dresden, Germany}

\author{Jeroen van den Brink}
\affiliation{Leibniz Institute for Solid State and Materials Research, Helmholtzstr. 20, D-01069 Dresden, Germany}

\author{Bernd B\"uchner}
\affiliation{Leibniz Institute for Solid State and Materials Research, Helmholtzstr. 20, D-01069 Dresden, Germany}
\affiliation{Institute for Solid State and Materials Physics, TU Dresden, D-01062 Dresden, Germany}
\affiliation{Würzburg-Dresden Cluster of Excellence ct.qmat, TU Dresden, 01062, Dresden, Germany}

\author{Vladislav Kataev}
\affiliation{Leibniz Institute for Solid State and Materials Research, Helmholtzstr. 20, D-01069 Dresden, Germany}

\author{Alexey Alfonsov}
\thanks{Email Address: a.alfonsov@ifw-dresden.de}
\affiliation{Leibniz Institute for Solid State and Materials Research, Helmholtzstr. 20, D-01069 Dresden, Germany}


\date{\today}

\begin{abstract}

The use of antiferromagnets in magnetoelectronic devices as counterparts of ferromagnets is a new, rapidly developing trend in spintronics that leverages antiferromagnetic (AFM) magnons for transmitting of spin currents. Van der Waals (vdW) antiferromagnets are particularly attractive in this respect as they possess tunable magnetic properties and can be easily integrated into spintronic devices. In this work we use electron spin resonance (ESR) spectroscopy to assess the potential of the vdW AFM compound \CCPS for magnonic applications by exploring the magnetic field ($H$) dependence of the spectrum of magnon excitations below its AFM ordering temperature $T_{\rm N} \approx 30$\,K and the correlated spin dynamics above $T_{\rm N}$. ESR reveals prominent ferromagnetic (FM) spin correlations that persist far above $T_{\rm N}$ suggesting an intrinsically two-dimensional character of the spin dynamics in \CCPS. Most interestingly, at $T < T_{\rm N}$, \CCPS features two non-degenerate, i.e., distinct in energy AFM magnon modes at $H = 0$ which can be tuned to the FM type of collective spin excitations with increasing $H$.
These remarkable properties are favorable for the induction and control of unidirectional spin current in \CCPS and suggest it as a new functional material for magnetoelectronics.

\end{abstract}

\maketitle

\section{Introduction}
Van der Waals (vdW) materials have garnered significant attention in solid state physics and materials research since the isolation of graphene~\cite{novoselov2004}. Remarkable success in exfoliation of this class of materials due to the lack of significant interlayer chemical bonds unlocks vast potential for applications in the fields of advanced electronics, optoelectronics, and spintronics~\cite{wang2012,chhowalla2013,mak2016,gong2019}. Furthermore, even in their bulk form, vdW materials often retain a quasi-two-dimensional (2D) character, making them particularly attractive for probing the low-dimensional physics while investigating bulk crystals.

Recent advances in synthesis and measurement techniques have reignited interest in magnetic vdW compounds, enabling a deeper exploration of magnetic anisotropies, spin excitations and spin-orbit coupling effects in the low-dimensional limit~\cite{lee2016,gong2017,wang2018,gong2019}. 
Besides fundamental aspects, such an interest is largely motivated by proposals to use vdW magnets as an element of magnon-spintronic devices which utilize spin currents carried by propagating magnons instead of electric current to transport and process information \cite{chumak2015,gong2019,Flebus2023,Manas2025}. Van der Waals magnets as a transmitting medium are beneficial in many respects, as they can be easily exfoliated down to a few- and monolayer limit and integrated into the device. Furthermore, they possess variable magnetic properties that can be tuned by chemical substitution or application of external stimuli, such as strain, light and electrical gating \cite{Burch2018,Khan2020,Yang2021}. Importantly, many of them are low-damping magnetic dielectrics which facilitates propagation of magnons over macroscopic length scales \cite{xing2019}. Also the circumstance that quite a few of these materials are  antiferromagnets is advantageous for their application in the new emergent field of antiferromagnetic spintronics promising ultra-fast and insensitive to disturbing magnetic fields signal processing superior to the ferromagnet-based devices \cite{Gomonay2018,Han2023}.        

\begin{figure*}[t]
	\centering
	\includegraphics[width=\linewidth]{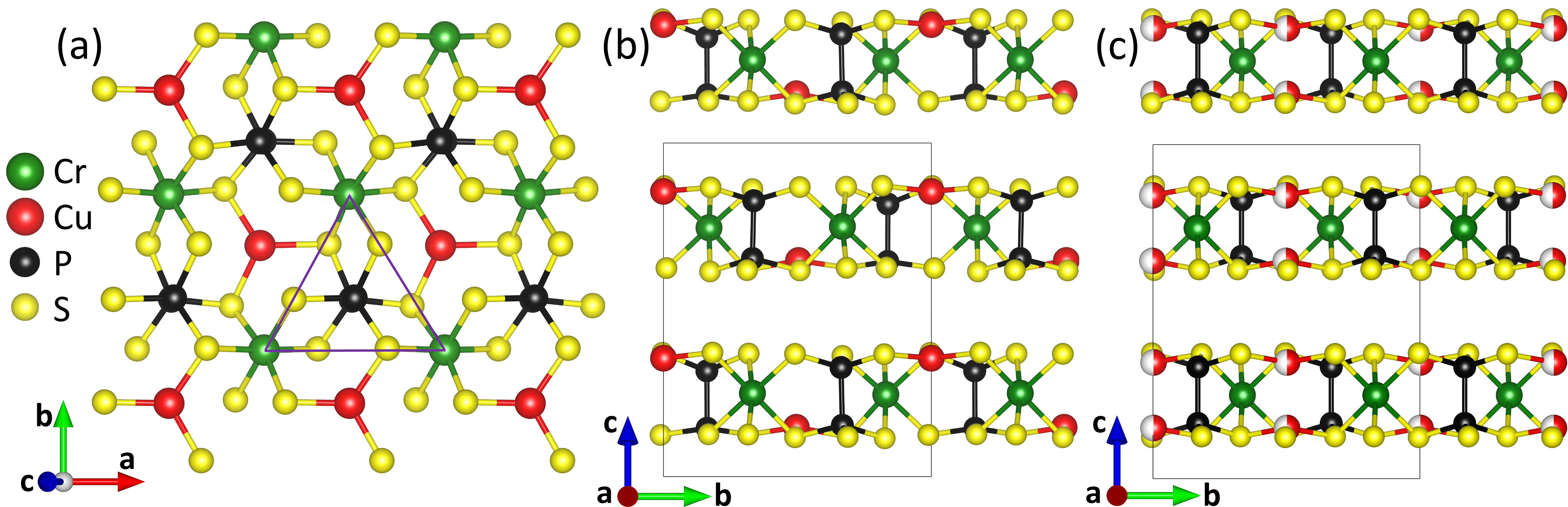}
	\caption{Crystal structure of \CCPS\ as determined in Ref.~\cite{maisonneuve1993}. (a) View normal to the $ab$-plane, i.e., along $c^*$-axis, showing the structure at $T = 64$\,K with Cr$^{3+}$ ions forming a 2D triangular magnetic lattice. (b) Quasi-2D layers stacked along the $c$-axis ($T = 64$\,K). (c) Same as (b), at high temperature ($T = 300$\,K) showing the distribution of Cu$^{1+}$ ions at two positions. The ratio of red and white colors denotes the site occupancy probability of Cu$^{1+}$ ions. Rectangle in (b) and (c) represents a unit cell.}
	\label{fig:Structure}
\end{figure*}

Within the broader class of these magnetic vdW materials, the transition metal thiophosphates of the form $AA^\prime$P$_2$S$_6$ ($A$ and $A^\prime$ refer to a transition metal ion), represent a particularly interesting subclass. In the absence of strong chemical bonding between crystallographic layers, the transition metal ions form 2D magnetic lattices in the layers weakly coupled in the third dimension. This material class offers high flexibility in selecting different transition metal ions at the $A$ and $A^\prime$ sites yielding a variety of magnetic properties. For instance, \FPS is an Ising-type antiferromagnet that preserves a long-range magnetic order at a significantly high transition temperature ($T_\mathrm{N}$) even in the monolayer limit~\cite{lee2016}. In contrast, in \NPS, which is an XXZ-type easy-plane antiferromagnet with a zigzag spin structure, magnetic order is completely suppressed in the 2D case~\cite{Kim2019}. Moreover, a remarkable  coupling of photoexcitations (excitons)
with antiferromagnetic order has been observed in thin \NPS single crystals, which may be relevant for developing antiferromagnet-based quantum information technologies~\cite{wang2021,hwangbo2021}. The \MPS, in which the spin order is of a N\'eel type with a weak easy-axis anisotropy, is found to host a long-distance magnon transport~\cite{xing2019} important for magnon-based spintronics and computing~\cite{chumak2015,Gomonay2018,Han2023,Flebus2023,Manas2025}.

The \CCPS vdW antiferromagnet crystallizing in the monoclinic $ C2/c $ space group \cite{maisonneuve1993,Colombet1982} is distinct from its above-mentioned counterparts in that the $A$ site in the individual \CCPS layer is occupied by the nonmagnetic Cu$^{1+}$ ($3d^{10}$) ions alternating with the magnetic Cr$^{3+}$ ($3d^3$, $S = 3/2$, $L = 0$) ions at the $A^\prime$ sites (Figure~\ref{fig:Structure}). While the spin-3/2 Cr$^{3+}$ ions reside in the center of the layer where they form a triangular lattice, the Cu$^{1+}$ ions are in the non-centrosymmetric positions. At high temperatures it is reported that there are two \cite{maisonneuve1993} (Figure~\ref{fig:Structure}(b)) or four \cite{Colombet1982} positions randomly occupied by Cu ions. However, at $\sim 144$\,K they structurally order at the alternating up and down sites giving rise to the antipolar (antiferroelectric) phase of \CCPS \cite{maisonneuve1993, lai2019,Susner2020,Cho2022}. The  Cr$^{3+}$ magnetic sublattice undergoes a transition to the magnetically ordered state at $T_{\rm N} \approx 30$\,K (Figure~\ref{fig:appendix_magn}(a) in Appendix~\ref{sec:appendix_magn} and Refs.~\cite{maisonneuve1993,maisonneuve1995,Kleemann2011,Susner2020}) where the Cr spins in each layer order ferromagnetically in the $ab$-plane with their corresponding moments aligned along the $b$-axis \cite{axes_avsb}. The neighboring layers are coupled antiferromagnetically  resulting in the A-type AFM spin structure \cite{maisonneuve1995}. Recent works indicate the presence of a magneto-electric coupling between the ordered Cu$^{1+}$ and Cr$^{3+}$ sublattices \cite{Cho2022,park2022}.

While electronic degrees of freedom of \CCPS have been addressed in bulk and thin-film samples \cite{lai2019,Cho2022,park2022, io2023}, studies of its magnetism -- essential for new functionalities -- largely focused on static properties \cite{maisonneuve1993,maisonneuve1995,Kleemann2011,Selter2023}, leaving the spin dynamics in the paramagnetic and AFM ordered phases practically unexplored. Although, Wang {\it et al.} \cite{wang2023} recently conducted a low-frequency antiferromagnetic resonance (AFMR) study capturing a small part of the spin excitation spectrum in the frequency range $\nu = 3 - 13$\,GHz, a comprehensive investigation of electron spin resonance (ESR) modes covering a wide range of temperatures, magnetic fields, excitation frequencies, and orientation dependences is still lacking. 
As reviewed in Refs.~\cite{Tang2023,Kataev2024}, typically, low-energy spin excitations in vdW magnets fall into the sub-THz frequency range, which is also the case for most of  the $AA^\prime$P$_2$S$_6$ family members \cite{Okuda1986,Kobets2009,Wyzula2022,mehlawat2022,senyk2023,abraham2023}. Thus, ESR spectroscopy an its related AFMR and ferromagnetic resonance (FMR) techniques covering a broad frequency domain are particularly needed to address fundamental questions of the spin dynamics in the quasi-2D spin systems, and  to assess the potential of vdW compounds  for applications in magnon-spintronics technologies. With this aim, we use these techniques   
to explore the low-energy spin dynamics in the short-range spin-correlated regime at temperatures above $T_{\rm N}$ and the spectrum of magnon excitations in the long-range AFM ordered state of a single crystal of \CCPS. Experimental studies are conducted in a broad temperature range 3 -- 300\,K at low magnetic fields $\mu_{0}H < 1$\,T with X-band ESR spectroscopy at the frequency $\nu = 9.56$\,GHz and in the high-field regime up to 16\,T in the frequency range 75 -- 330\,GHz  with high-frequency electron spin resonance (HF-ESR) spectroscopy. Moreover, we perform DFT calculations in order to get deeper insights into the nature of the magnetic interactions and anisotropies in this system. First, we find that in the paramagnetic regime, signatures of the in-plane FM correlations between the Cr spins can be seen in the ESR response far above $T_{\rm N}$, becoming particularly pronounced in strong magnetic fields and evidencing an inherent quasi-2D character of the magnetism of bulk \CCPS. Second, in the AFM ordered state below $T_{\rm N}$ we discover new, previously unreported, resonance modes and study them in the $\nu - H$ parameter domain for different directions of the applied magnetic field. The analysis of the data using linear spin wave theory (LSWT) enables a precise quantification of the magnetic anisotropy constants which determine the in-plane orientation of the Cr spins with the preferable direction along the $b$-axis. Additionally, this analysis directly yields the strength of the inter-plane magnetic coupling necessary for stabilization of the A-type AFM order and enables us to identify two magnon excitation  modes with distinct energies in the zero-field limit. Importantly, the calculated parameters of the spin Hamiltonian using density functional theory (DFT) with generalized gradient approximation (GGA+U) show a very good agreement with those estimated from the measurements. Third, we observe a remarkable effect of the field-tuning of the character of the resonance modes from the AFM type at $\mu_{0}H < 6-8$\,T to the FM type at stronger fields. This unusual effect can be explained by considering the closeness of the  scales of the inter-plane exchange and  the magnetic anisotropy energies. The non-degeneracy of the magnon modes at zero field, which is required to excite the unidirectional magnonic spin current in the material, as well as a controllable tuning of the type of the spin excitations by the magnetic field appear as promising functionalities of \CCPS in terms of its application in magnonic devices \cite{Gomonay2018,Han2023,Flebus2023}. Finally, no indications of the antiferroelectric order in the Cu$^{1+}$ sublattice is found in the ESR measurables within the experimental error bars, suggesting the weakness of the cross-coupling between the electrical and magnetic subsystems of \CCPS similar to other type-I multiferroics.

\section{Results}

\begin{figure}[t]
	\centering
	\includegraphics[width=1.0\linewidth]{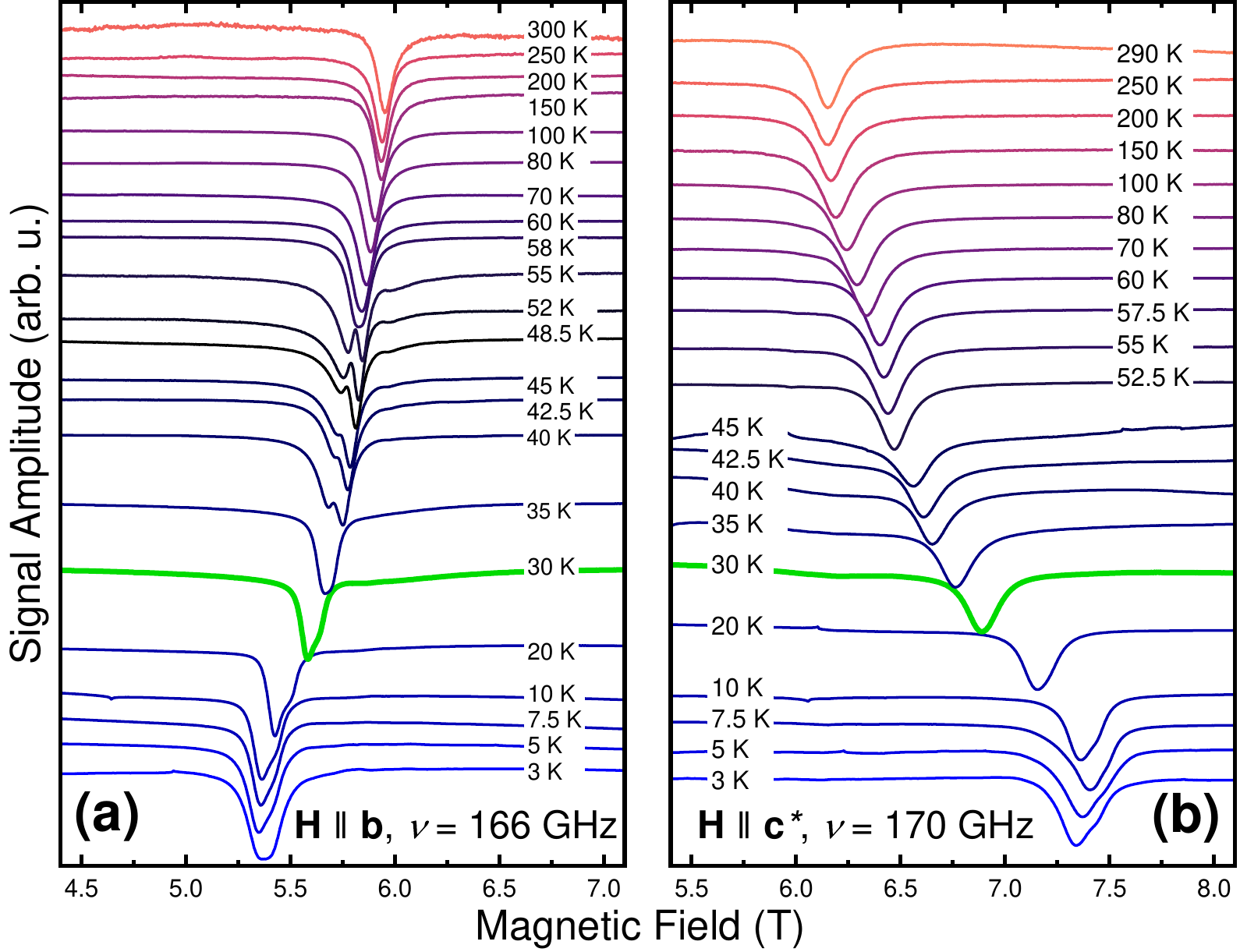}
	\caption{HF-ESR spectra of \CCPS at various temperatures for (a) \HIIb configuration at 166 GHz and for (b) \HIIc configuration at 170 GHz. The spectral shape was corrected to eliminate the dispersive component of the detected signal as explained in \cite{pal2024}. The spectra are also normalized and shifted vertically for clarity.}
	\label{fig:Tdep_HF}
\end{figure}

\begin{figure*}[!t]
	\centering
	\includegraphics[width=\linewidth]{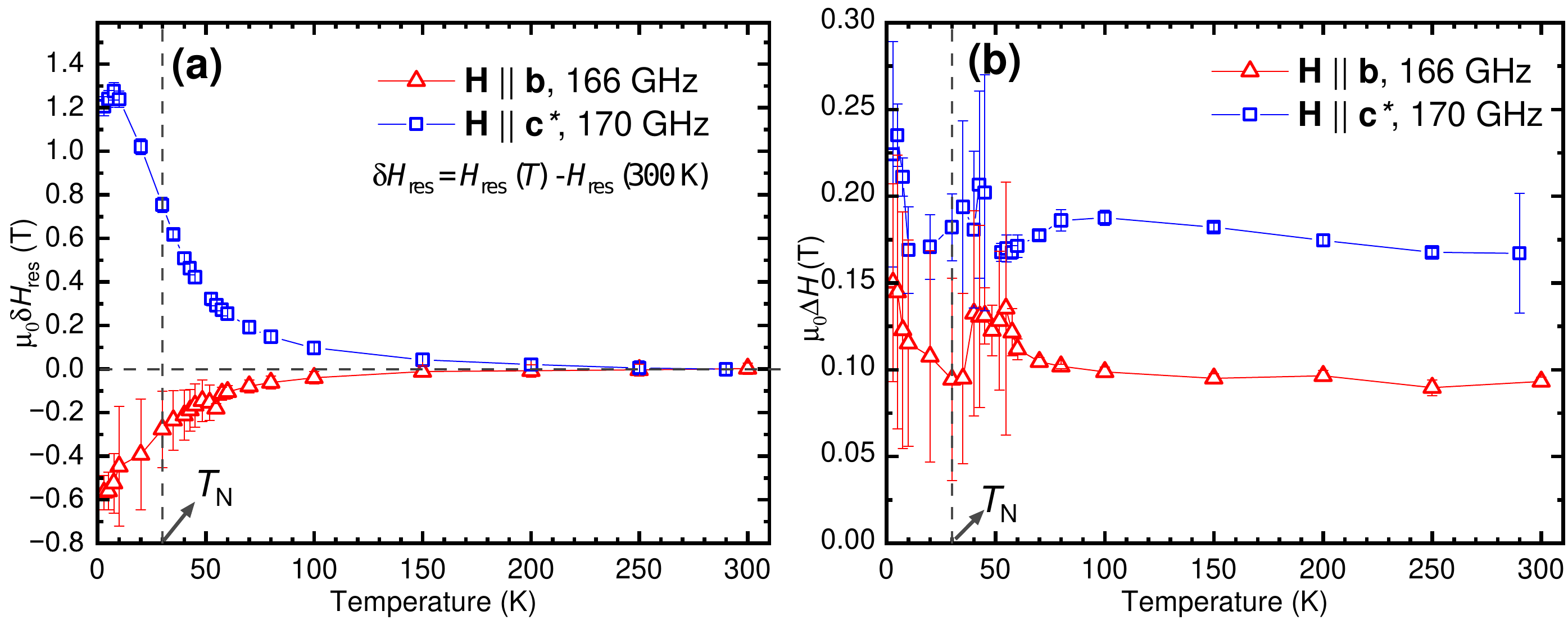}
	\caption{(a) Temperature dependence of the shift $\delta H (T) = H_{res} (T) - H_{res} (300\,{\rm K})$ of the resonance field from the paramagnetic position denoted by the horizontal dashed gray line for both orientations. (b) Evolution of the linewidth $\Delta H$ as a function of temperature. The vertical dashed gray line on both plots indicates the AFM transition temperature $T_{\rm N}$ of \CCPS. Error bars less than the symbol size are omitted. Lines connecting data points are guides for the eye.}
	\label{fig:Hres_LW}
\end{figure*}

\subsection{Temperature dependence of HF-ESR}

The temperature dependent evolution of the HF-ESR spectra was measured for a magnetic field configuration \HIIb at 166 GHz and \HIIc at 170 GHz as depicted in Figure \ref{fig:Tdep_HF}. Note, that ${\bf c^*}$ is the direction perpendicular to the $ab$-plane (Figure~\ref{fig:Structure}(a)). Upon cooling the sample from room temperature a significant shift in the spectral position is noticeable already far above $T_\mathrm{N}$ = 30\,K. To illustrate this, the temperature-dependent shift [$\delta H (T) = H_{res} (T) - H_{res} (300\,{\rm K})$] of the experimentally obtained resonance position, $H_{res} (T)$ from the expected resonance position in the paramagnetic phase $H_{res}$ (300\,K) is plotted in Figure \ref{fig:Hres_LW}(a). $H_{res}(300\,\rm K)$ is calculated using the equation which describes the paramagnetic resonance condition: 
\begin{equation}
	h\nu = g\mu_{B}\mu_{0}H_{res}(300\,{\rm K})\ .
	\label{eq:FDep_300K}
\end{equation}
Here, $h\nu$ represents the energy of the microwaves, $\mu_{B}$ is the Bohr magneton and $\mu_{0}$ denotes the permeability of free space. $ g $ signifies the  $ g $-factor of the Cr spins obtained from the frequency dependence at 300\,K for \HIIb and \HIIc configurations, as described in Section~\ref{sec:FDep_300K}. The onset of the deviation in the $\delta H (T)$ plot is seen at about 200 -- 250\,K. The splitting of the ESR lines below 60\,K (Figure~\ref{fig:Tdep_HF}(a)) for the \HIIb configuration is likely  an instrumental effect and is accounted for in the enlarged error bars in Figure~\ref{fig:Hres_LW}~\cite{peak_splitting}. For this reason, particularly the temperature dependence of the linewidth $\Delta H (T)$ (Figure~\ref{fig:Hres_LW}(b)) shows significant scattering of the data points below 60\,K.

\subsection{Frequency dependence in the paramagnetic phase}
\label{sec:FDep_300K}

\begin{figure}[!t]
	\centering
	\includegraphics[width=1.0\linewidth]{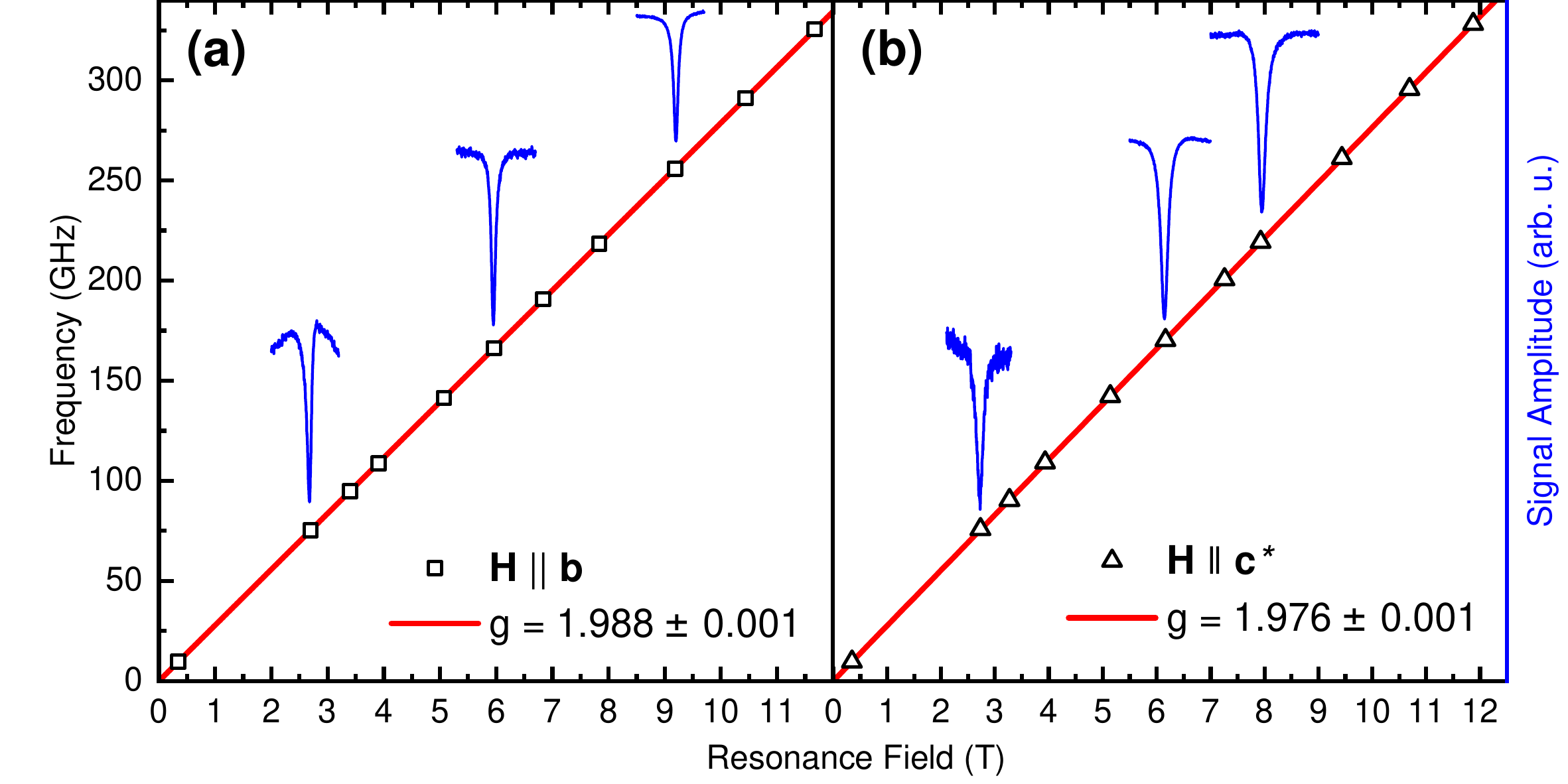}
	\caption{Frequency dependence $\nu(H_{\rm res}$) at 300\,K for (a) \HIIb  and (b) \HIIc shown by open squares and triangles, respectively. The solid red line is the fit to the paramagnetic resonance condition (Equation~(\ref{eq:FDep_300K})). Right vertical axis: representative ESR signals at different frequencies normalized and vertically shifted above the corresponding $\nu$ vs $H_{res}$ data points. The $g$-factor values are summarized in the table in Appendix~\ref{sec:abc_prop}}
	\label{fig:FDep_300K}
\end{figure}

To determine the $g$-factor anisotropy in the paramagnetic phase, the dependence of the excitation frequency ($\nu$) on the resonance field ($H_{res}$) of the ESR signal was measured for both \HIIb and \HIIc orientations at room temperature (Figure~\ref{fig:FDep_300K}). $\nu$ scales linearly with $H_{res}$ according to the simple paramagnetic resonance condition described by Equation~(\ref{eq:FDep_300K}), as expected. The respective fit yields the slightly anisotropic $g$-factor values $ g_{\parallel c*} = 1.976\pm 0.001$ and $ g_{\parallel b} = 1.988\pm 0.001$, close to the spin-only $g$-factor of 2. Since the Cr$^{3+}$ ion in the high-spin state $S = 3/2$ has zero orbital angular momentum in first order, the $g$ values are expected to be close to $g = 2$ and to have a possible slight anisotropy arising from the second order spin-orbit coupling effect \cite{Abragam2012}. 

\subsection{Frequency dependence in the AFM ordered phase}
\label{sec:FDep_3K}

\begin{figure*}[!t]
	\centering
	\includegraphics[width=\linewidth]{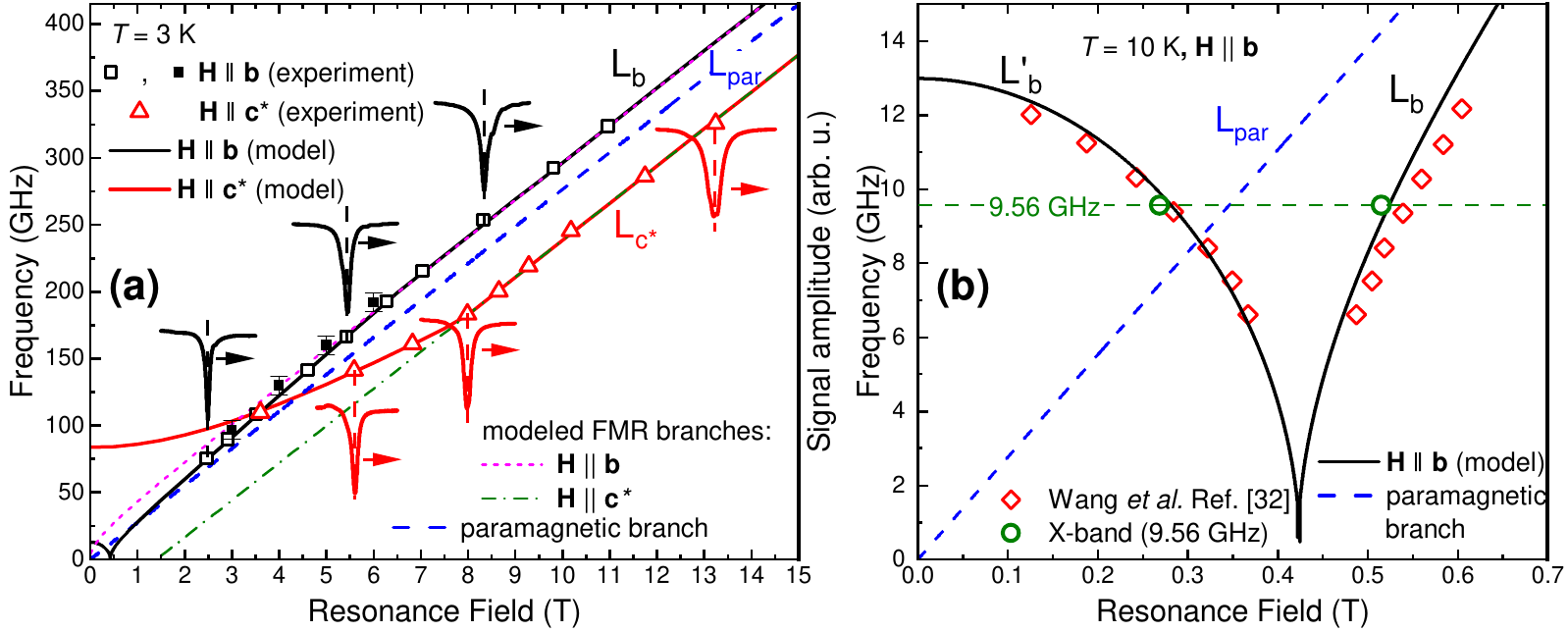}
	\caption{(a) $\nu(H_{\rm res})$ dependence of the resonance modes (branches) at 3\,K for field configurations \HIIb and \HIIc (open symbols). Solid squares represent the results of the ESR measurements in the frequency domain for \HIIb configuration (see Appendix~\ref{sec:FreqDom}). Solid black and red lines denoted as L$_{\rm b}$ and L$_{\rm c^\ast}$ branches are the result of the modeling of the experimental data based on Hamiltonian~(\ref{Hamil}). Dashed blue line represents the paramagnetic branch L$_{\rm par}$ described by Equation~(\ref{eq:FDep_300K}). The dot-dashed line in green and the short-dashed line in magenta are modeled FMR branches calculated by setting $A_{\rm ex} = 0$ in Hamiltonian~(\ref{Hamil}) (for details see the text). Right vertical axis: exemplary spectra (normalized and shifted vertically) for both field orientations at selected frequencies. (b) Zoomed-in low-field part of the $\nu(H_{\rm res})$ diagram in panel (a). Diamonds are the data from Wang {\it et al.}~\cite{wang2023} for the magnetic field lying along easy axis \cite{axes_avsb} at $T = 10$\,K, open circles are the resonance fields for \HIIb of the signals measured at the X-band frequency of 9.56\,GHz which is indicated by the horizontal dashed line (see Section~\ref{sec:X-band}). The notation of the solid black and the dashed blue lines is the same as in panel (a) (for details see the text).}
	\label{fig:FDep_3K}
\end{figure*}

Low-temperature AFMR modes (the uniform spin wave modes in the wave vector ${\bf q} \rightarrow 0$ limit) were explored by HF-ESR in the frequency range $\nu = 75 - 330$\,GHz. Their $\nu(H_{\rm res})$ dependence (resonance branches) for \HIIb and \HIIc field geometries is presented in Figure \ref{fig:FDep_3K}(a). 

In contrast to the ESR signals in the paramagnetic regime that follow a simple gapless linear $\nu(H_{\rm res})$ dependence defined by Equation~(\ref{eq:FDep_300K}) (branch L$_{\rm par}$ in Figure~\ref{fig:FDep_3K}), the AFMR modes exhibit a non-linear behavior strongly dependent on the direction of the applied magnetic field. In particular the mode for \HIIc tends to have a finite frequency offset of $\sim 80$\,GHz if its dependence is approximated to zero field. Such a newly discovered gap for the spin excitations is also confirmed by the experiments performed in the frequency domain, i.e., by sweeping the microwave frequency while keeping the magnetic field at a constant value (see Appendix~\ref{sec:FreqDom}).

To analyze the $\nu(H_{\rm res})$ dependence of the observed resonance modes in \CCPS we applied the LSWT with the second quantization formalism \cite{Turov, Holstein1940}. Our approach for using this model is detailed in Ref.~\cite{alfonsov2021}. For this purpose the A-type AFM ordered spin lattice of \CCPS can be represented by ferromagnetic layers in the $ab$-plane coupled antiferromagnetically along the $c^\ast$-axis. This can be described using the two-sublattice model with the staggered sublattice magnetizations  $\boldsymbol{M_1}$ and $\boldsymbol{M_2}$ such that $\boldsymbol{M_1}^2 = \boldsymbol{M_2}^2 = (M_0)^2 = (M_\mathrm{S}/2)^2$, where $M_\mathrm{S}$ is the saturation magnetization. The respective phenomenological Hamiltonian used in this model takes the following form:

\begin{align}
	\label{Hamil}
	\mathscr{H} = \ & A_\mathrm{ex} \frac{(\boldsymbol{M_1 M_2})}{M_0^2} + K_\mathrm{uniax} \frac{{M_{1_z}}^2 + {M_{2_z}}^2}{M_0^2} \nonumber \\
	+ \ & \frac{K_\mathrm{biax}}{2} \frac{(M_{1_x}^2 - M_{1_y}^2) + (M_{2_x}^2 - M_{2_y}^2)}{M_0^2} \nonumber \\
	- \ & (\boldsymbol{\textbf{H} M_1}) - (\boldsymbol{\textbf{H} M_2})\, .
\end{align}
Here, the first term represents the mean-field antiferromagnetic exchange energy density characterized by the constant $A_{\rm ex}$. The second and the third term define the uniaxial and the biaxial components of the magnetocrystalline anisotropy energy densities with the respective constants $K_{\rm uniax}$ and $K_{\rm biax}$. The sign of the former dictates either the in-plane or the out-of-plane orientation of the sublattice magnetizations whereas the latter defines the anisotropy in the plane. Lastly, the fourth and fifth terms correspond to the Zeeman interaction for both sublattice magnetizations. The LSWT numerical modeling of the $\nu(H_{\rm res})$ dependences in Figure~\ref{fig:FDep_3K}(a) based on Hamiltonian~(\ref{Hamil}) was carried out with $A_\mathrm{ex}$, $K_\mathrm{uniax}$ and $K_\mathrm{biax}$ as free parameters. The saturation magnetization for Cr$^{3+}$ ion $M_{\rm S} = gS\mu_{\rm B}$ was calculated with the spin value $S = 3/2$ and the average $g$-factor of 1.98 obtained from the high temperature HF-ESR measurements (Figure~\ref{fig:FDep_300K}) yielding $M_\mathrm{S}$ = 143\,erg/(G cm$^3$). The best agreement with our HF-ESR experimental data was achieved with the parameters of Hamiltonian (\ref{Hamil}) that are summarized in Table~\ref{tab:parameters}. The modeled resonance branches $\nu(H_{\rm res})$ are depicted in Figure~\ref{fig:FDep_3K} by solid lines. We note that a contribution of the shape anisotropy due to the platelike  form of the studied crystal is negligible because of the small magnetization of the sample at $T \ll T_{\rm N}$ and therefore it does not affect the results of the modeling (see Appendix~\ref{Sec:Shape_anisotropy}).

\begin{table*}[t!]
	\centering
	\begin{tabular}{|l|c|c|c|c|c|c|}
		
		\multicolumn{1}{c}{} & \multicolumn{3}{c}{Experimental} & \multicolumn{3}{c}{Theoretical}\\ \hline
		& erg cm$^{-3}$ & J m$^{-3}$ & meV/Cr$^{3+}$ & erg cm$^{-3}$ & J m$^{-3}$ & meV/Cr$^{3+}$ \\ \hline
		
		$A_\mathrm{ex}$ & $23 \times 10^5$ & $23 \times 10^4$ & $0.280$ 
		& $25 \times 10^5$ & $25 \times 10^4$ & $0.305$ \\ \hline
		
		$K_\mathrm{uniax}$ & $5 \times 10^5$ & $5 \times 10^4$ & $0.061$
		& $9 \times 10^5$ & $9 \times 10^4$ & $0.11$ \\ \hline
		
		$K_\mathrm{biax}$ & $0.1 \times 10^5$ & $0.1 \times 10^4$ & 1.2 $\times 10^{-3}$   & $0.24 \times 10^5$ & $0.24 \times 10^4$ & $3 \times 10^{-3}$ \\ \hline
	\end{tabular}
	\caption{Ground state energy parameters as obtained from the LSWT modeling described by Equation~\ref{Hamil} (Experimental) and by the DFT calculations (Theoretical) (Section~\ref{sec:DFT_results}).}
	\label{tab:parameters}
\end{table*}

\subsection{DFT calculations}
\label{sec:DFT_results}

The unit cell of \CCPS\ contains two magnetic layers oriented parallel to the $ab$ plane (Figure~\ref{fig:Structure}(b,c)). Both Cr and Cu atoms are coordinated by six S atoms, forming a honeycomb lattice with P-P dimers in the voids. However, the octahedral coordination of monovalent Cu is unstable, and the Cu atoms are significantly shifted along the $c$ axis. At high temperatures, these displacements are not long-range ordered, and hence according to \cite{maisonneuve1993} the crystal structure features Cu sites with 50\% occupancy (Figure~\ref{fig:Structure}(c)). In the low-temperature structure, Cu atoms are ordered, yet the experimental information on their exact arrangement in the crystal structure is scarce. Based on neutron powder diffraction data, Maisonneuve {\it et al.} \cite{maisonneuve1993} suggested a model with an antiferroelectric order described by the space group $Pc$ (Figure~\ref{fig:Structure}(b)). However, other arrangements (including other antiferroelectric arrangements) are possible. To study this in more detail, we started with the high-temperature $C2/c$ structure and constructed five different ordered configurations shown in Figure~\ref{fig:CuConfiguration}(b-f) of Appendix~\ref{Sec:DFT_appendix}: one ferroelectric, one ferrielectric, and three antiferroelectric. By comparing the total energies of these ordered configurations, we found that the lowest-energy configuration is antiferroelectric, as depicted in Figure~\ref{fig:Structure}(b) and Figure~\ref{fig:CuConfiguration}(b), yet in contrast to Ref.~\cite{maisonneuve1993} it is described by the space group $P2_1$.

After the optimization of this structure depicted in Figure~\ref{fig:Structure}(b) we used its parameters to calculate the interlayer exchange and magnetic anisotropy constants. The crystal structure of \CCPS\ features many possible interlayer connections with similar $d_{\text{Cr-Cr}}$ distances that can facilitate magnetic exchange. An accurate estimation of all these individual magnetic exchanges would require total energy calculations of very large supercells. Instead, we consider here the effective interlayer exchange constant $A^{inter}_{cal}$, which corresponds to the energy cost of flipping all magnetic moments in one magnetic layer. Importantly, this quantity can be directly compared with the exchange constant obtained from experiment. We estimate this energy cost of flipping all magnetic moments as the difference between the total energies of the ferromagnetic and the A-type antiferromagnetic configurations. 
The calculated $A^{inter}_{cal}$ is 25 $\times$ 10$^4$ J/m$^3$ (0.305 meV/f.u). To calculate the uniaxial anisotropy constant $K^{uniax}_{cal}$ and biaxial anisotropy constant $K^{biax}_{cal}$, we performed spin-polarized relativistic GGA+$U$ calculations, with the spin moments along \HIIa, \HIIb and \HIIc directions. 
The results showed that the easy axis is the $b$-axis and that the hard axis is the $c^\ast$-axis, and the calculated $K^{uniax}_{cal}$ = 9 $\times$ 10$^4$ J/m$^3$ and $K^{biax}_{cal}$ = 2.4 $\times$ 10$^3$ J/m$^3$.

As can be seen in Table~\ref{tab:parameters}, the results of the calculations match well with the estimations of the Hamiltonian parameters from the experiment.

\subsection{X-Band ESR}
\label{sec:X-band}

\begin{figure}[!t]
	\centering
	\includegraphics[width=1.0\linewidth]{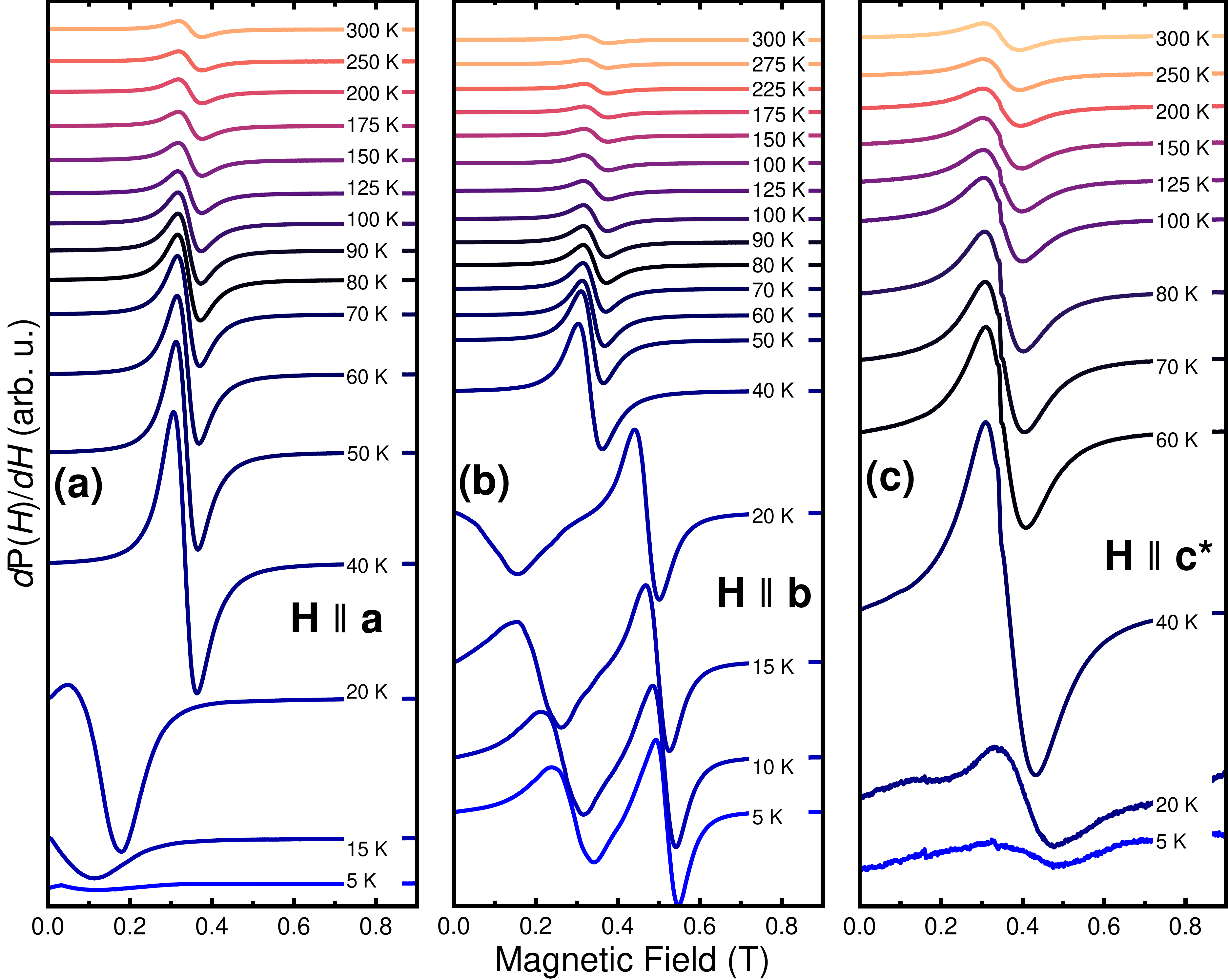}
	\caption{Temperature evolution of the X-band ($\nu = 9.56$\,GHz) ESR spectra (field derivatives of the microwave absorption $dP(H)/dH$) for (a) \HIIa, (b) \HIIb, and (c) \HIIc field geometry.}
	\label{fig:Tdep_Spectra}
\end{figure}

\begin{figure*}[!t]
	\includegraphics[width=\linewidth]{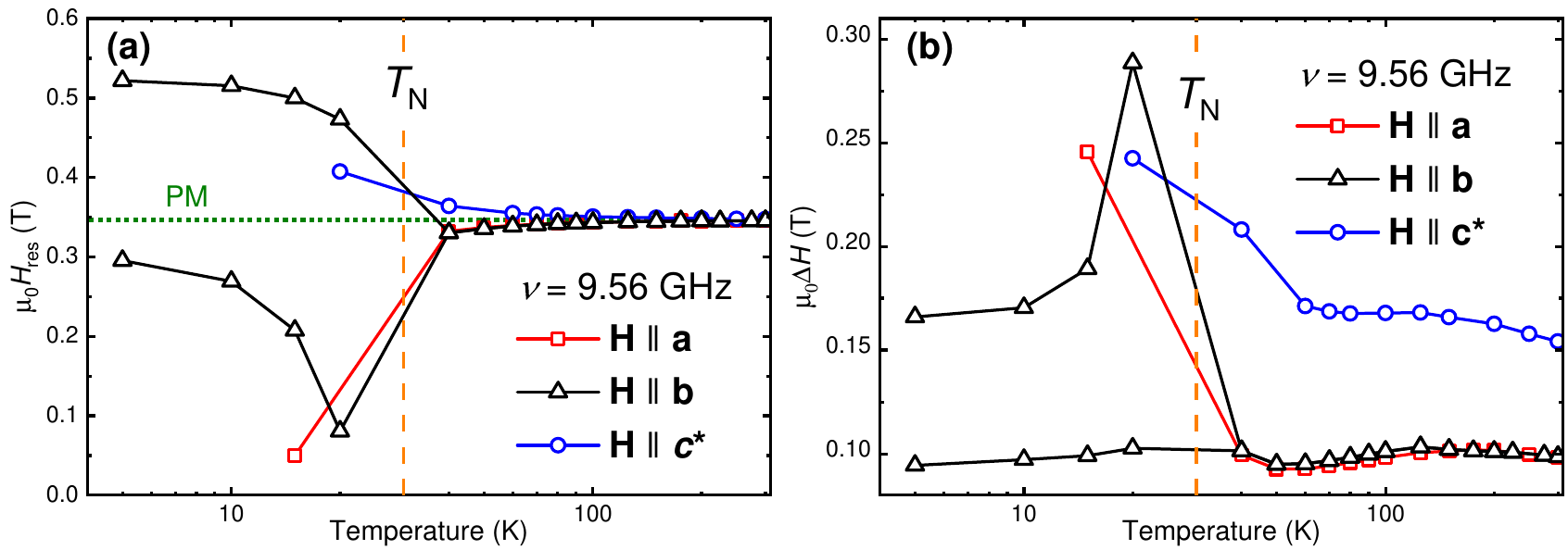}
	\caption{Temperature dependence of the resonance field $H_{\rm res}(T)$ (a) and of the linewidth $\Delta H(T)$ (b) for \HIIa, \HIIb and \HIIc. The lines connecting symbols in (a) and (b) are guides to the eye. The vertical dashed line in both plots indicates the AFM transition temperature $T_{\rm N}$ and the horizontal green dotted line PM in (a) represents the paramagnetic resonance position of the ESR signal defined by Equation~(\ref{eq:FDep_300K}) using the average values of $g_{\parallel \mathrm{c^{\ast}}}$ and $g_{\parallel \mathrm{b}}$ at 300\,K. Note the logarithmic temperature scale in both plots.}
	\label{fig:Tdep_XBand}
\end{figure*}

The temperature dependence of the ESR spectra  was measured at the X-band frequency of 9.56\,GHz for the magnetic field applied along the three crystallographic directions {\bf a}, {\bf b} and ${\bf c^\ast}$, and their angular dependence was studied for the rotation of the applied field in the $ab$- and $bc^\ast$-planes at selected temperatures. The temperature evolution of the ESR spectra is presented in Figure~\ref{fig:Tdep_Spectra}. In the paramagnetic state above $T_{\rm N}$ the spectrum consists of a single Lorentzian absorption derivative line $dP(H)/dH$. For \HIIb the signal splits into two components at $T < T_{\rm N}$. The corresponding temperature dependence of the resonance field $H_{\rm res}(T)$ is shown in Figure~\ref{fig:Tdep_XBand}(a). The position of these split lines is in a good correspondence with the data in Ref.~\cite{wang2023} and with the LSWT analysis of the HF-ESR results (see Figure~\ref{fig:FDep_3K}(b)). For the other two directions, the X-band ESR signal shifts out of the experimental field range or looses the intensity at $T \lesssim  T_{\rm N}$. The onset of the deviation of $H_\mathrm{res}(T)$ from the expected paramagnetic position, which is indicated by the green dashed line in Figure~\ref{fig:Tdep_XBand}(a), can be clearly discerned at $T\lesssim 125$\,K while it is appreciable already at $ T \sim 200 - 250$\,K in the HF-ESR data (Figure~\ref{fig:Hres_LW}(a)). The temperature dependence of the linewidth experiences a  narrowing effect below $\sim 150$\,K followed by a critical broadening by approaching $T_{\rm N}$ (Figure~\ref{fig:Tdep_XBand}(b)).

\begin{figure*}[!t]
	\includegraphics[width=\linewidth]{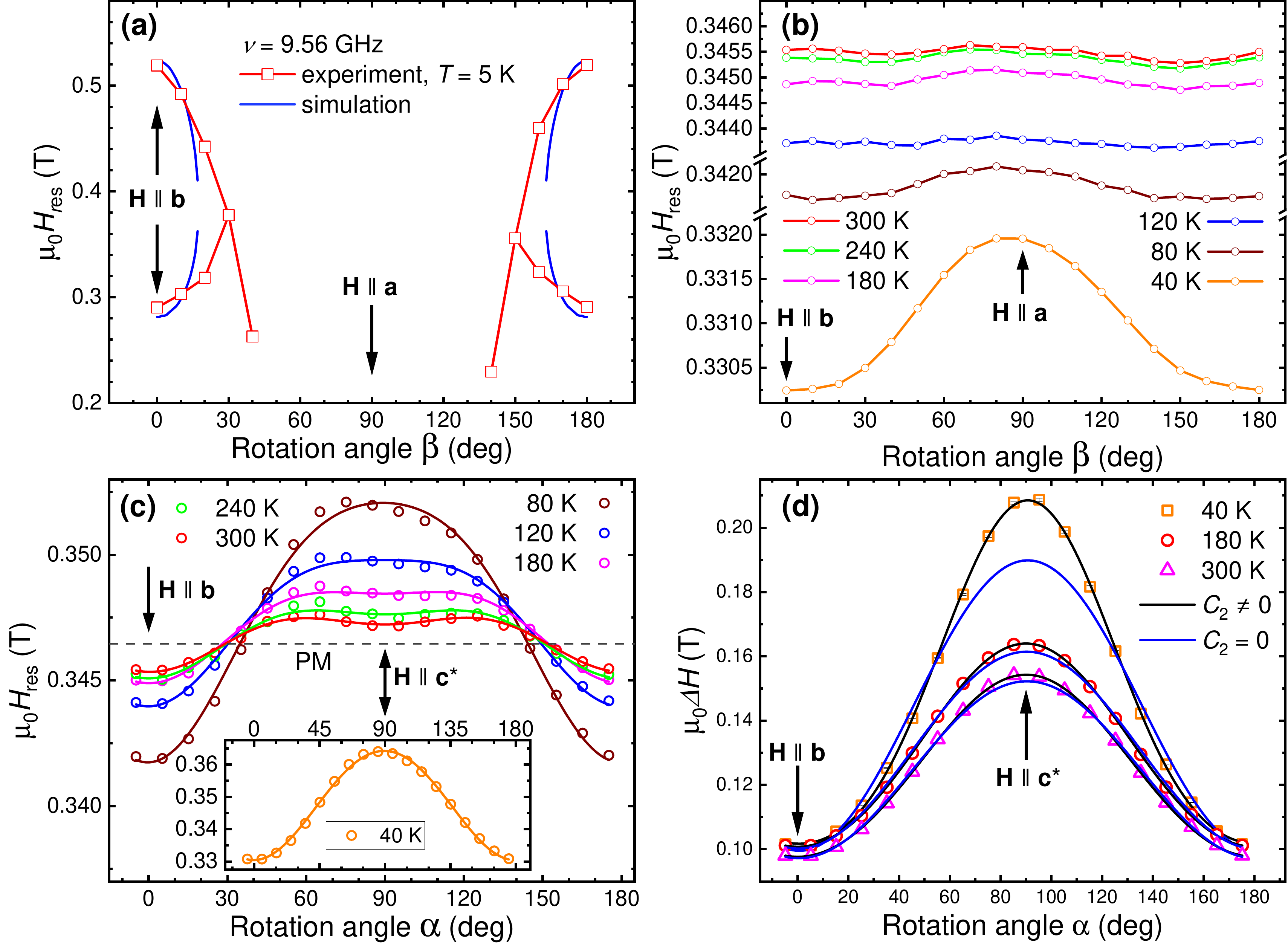}
	\caption{Angular dependence of the  resonance field $H_{\rm res}$ for rotation of the magnetic field vector in the $ab$-plane at (a) $T = 5$\,K and (b) in the temperature range $T =40 - 300$\,K, and (c) in the $bc^\ast$-plane at  $T =40 - 300$\,K. Blue solid lines in (a) are fits using the LSWT approach (see the text). Solid lines in (c) are the fits to the experimental data at each temperature to Equation~(\ref{eq:Hres}) as described in Appendix~\ref{sec:Hres_angular}. Horizontal dashed line in (c) represents the resonance field defined as the mean of $H_{\rm res}(\alpha)$ at 300\,K ($H_{\rm res}^{\rm mean}$ in Appendix~\ref{sec:Hres_angular}). (d) Angular dependence of the linewidth $\Delta (H)$ in the $bc^\ast$-plane at $T =$ 40, 180 and 300\,K (open symbols). The blue and black solid lines are fits with Equation~(\ref{eq:LW}) to the data with coefficients $C_2=0$ and $C_2\neq0$ respectively, as explained in the text. }
	\label{fig:Adep_XBand}
\end{figure*}

The rotation of the magnetic field vector ${\bf H}$ by the angle $\beta$ from the $b$- to the $a$-axis in the AFM ordered phase at $T = 5\,K \ll T_{\rm N}$ causes a rapid merging of the split resonance lines (the low-$T$ spectrum in Figure~\ref{fig:Tdep_Spectra}(b)) into a single line. By further rotation of ${\bf H}$ this merged line is shifted out of the available magnetic field range (Figure~\ref{fig:Adep_XBand}(a)). Notably, the angular variation of the split lines can be well modeled with the LSWT approach based on Hamiltonian~(\ref{Hamil}) with the parameters obtained in the analysis of the HF-ESR data (blue solid lines in Figure~\ref{fig:Adep_XBand}(a)). With increasing the temperature above $T_{\rm N}$, the $H_{\rm res}(\beta)$ dependence of the ESR signal is uniaxial and practically vanishes above $T\sim 80$\,K (Figure~\ref{fig:Adep_XBand}(b)).

The evolution of the angular dependence of $H_{\rm res}$ at $T > T_{\rm N}$ is different  when the field vector rotates by the angle $\alpha$ in the $bc^\ast$-plane. The amplitude of $H_{\rm res}(\alpha)$ is much larger than that of the $H_{\rm res}(\beta)$ dependence and follows a simple 180$\degree$ periodicity (Figure~\ref{fig:Adep_XBand}(c)(inset)) due to the uniaxial anisotropic internal field persisting with decreasing magnitude well above $T_{\rm N}$ (Figure~\ref{fig:Tdep_XBand}(a)). With further increase of the temperature towards 300\,K the amplitude of the variation of $H_{\rm res}(\alpha)$ approaches that determined by the anisotropy of the $g$-factors (Section~\ref{sec:FDep_300K}). Interestingly, in this temperature regime $H_{\rm res}(\alpha)$ develops a local minimum at $\alpha = 90\degree$ (\HIIc) suggesting a continuous deviation of the symmetry of the magnetic anisotropy from two-fold uniaxial-- towards a more complex four-fold symmetry case (see Appendix~\ref{sec:Hres_angular}).

Finally, the angular dependence of the ESR linewidth $\Delta H(\alpha)$ for ${\bf H}$ rotating in the $bc^\ast$-plane in the first approximation reveals a uniaxial character at all temperatures above $T_{\rm N}$ with the amplitude decreasing with raising temperature (Figure~\ref{fig:Adep_XBand}(d)). To verify this, we use a phenomenological equation for the angular dependence of the linewidth defined as \cite{senyk2023}:
\begin{align}
	\label{eq:LW}
	\mu_0 \Delta H = C_1 (\cos^2 \theta + 1) + C_2 (3\cos^2 \theta - 1)^2 + C_3 ,
\end{align}
where $\theta = 90^\circ - \alpha$ is the angle between the applied field and the $\rm c^\ast$-axis. The first term describes the angular dependence of the linewidth typical for three-dimensional (3D) spin systems with a significant anisotropic coupling, the second term represents the contribution due to the 2D-type spin-spin correlations, and the third term is the angle-independent part of the linewidth \cite{Richards1974,Benner1978,Benner1990,senyk2023, Moro2022}. At all temperatures, this function provides a reasonably good fit to the angular dependence. While at high temperatures the contribution of the 2D term is minor, it becomes significant at lower temperatures. Indeed, setting $C_2$ to zero yields in this temperature regime a significant deviation of the fit from the experimental dependence (Figure~\ref{fig:Adep_XBand}(d)) suggesting the prominence of the 2D-type dynamics in the spin system at lower temperatures by approaching $T_{\rm N}$.

\section{Discussion}
\subsection{Spin dynamics at $T > T_{\rm N}$}
\label{sec:spin_dynamics}

In conventional 3D magnets the stabilization of magnetic order is directly related with the emergence of the static internal magnetic field, which causes a shift of the ESR line from the paramagnetic position at the ordering temperature. Typically, in quasi-2D spin systems the in-plane dynamic spin-spin correlations are already prominent far above this temperature since low dimensionality inhibits the establishment of 3D long-range order due to the weakness of the inter-plane magnetic interactions. Such slowly fluctuating, persisting short-range order appears static on a fast, nanosecond ESR timescale and, therefore, causes a shift of the resonance field $\delta H_{\rm res}$ and affects the anisotropy of the linewidth \cite{Benner1990}. Consequently, the temperature development of this short-range order, and its respective transformation into a long-range ordered state, as seen at the ESR timescale, is usually gradual, i.e., without any anomaly at the ordering temperature \cite{Kataev2024, senyk2023,abraham2023,pal2024,alfonsov2021,zeisner2019}.

Remarkably, in the case of \CCPS the magnitude of the shift of the ESR line at temperatures much higher than $T_{\rm N}$ is significantly larger for HF-ESR (Figure~\ref{fig:Hres_LW}(a)) as compared to X-band ESR (Figure~\ref{fig:Tdep_XBand}(a)). This indicates the magnetic field induced enhancement of spin correlations which is expected if they are of the FM type. Such a field enhancement was indeed observed, e.g., in the ESR experiments on the vdW ferromagnet Cr$_2$Ge$_2$Te$_6$ \cite{zeisner2019}. Thus, current observations support the proposed hierarchy of magnetic interactions in \CCPS with strong FM coupling between the Cr spins in the planes contrasting with much weaker AFM interplane interaction \cite{Colombet1982,maisonneuve1995,park2022}. The sign of $\delta H_{\rm res}$ depends on the type of magnetic anisotropy which is maintained in the short-range ordered state above $T_{\rm N}$. If it is out-of-plane (easy axis) then $\delta H_{\rm res}$ is negative for $\bf H$ normal to the plane and positive for the in-plane field geometry, like it is found for Cr$_2$Ge$_2$Te$_6$ \cite{zeisner2019}. It is {\it vice versa} for the in-plane (easy-plane) anisotropy as evidenced by the positive $\delta H_{\rm res}$ shift for \HIIc and the negative shift for \HIIb in the case of \CCPS (Figure~\ref{fig:Hres_LW}(a)). The resulting angular dependence of the resonance field $H_{\rm res}(\alpha,\beta)$ in the spin-spin correlated regime above $T_{\rm N}$   reflects the uniaxial character of magnetic anisotropies in \CCPS for rotations from the in-plane easy $b$-axis to the hard in-plane $a$-axis and back (Figure~\ref{fig:Adep_XBand}(b)), and for rotations from the easy $ab$-plane to the hard out-of-plane $c^\ast$-axis and back (Figure~\ref{fig:Adep_XBand}(c)).

One particular signature of the 2D spin system is a specific angular dependence of the ESR linewidth of the type $\Delta H \propto (3 \cos^{2}(\theta)-1)^2$ (Equation~(\ref{eq:LW})) originating in the slow time decay of the spin self-correlation function \cite{Richards1974,Benner1978,Benner1990}. This type of dependence is predicted for both FM and AFM intraplane interactions. In the quasi-2D magnets with antiferromagnetic interaction between the  spins in the layers the onset of  a 3D long-range order causes a shortening of the time decay of the spin correlation function. It results in a change of the character of the angular dependence to $\Delta H \propto (\cos^{2}(\theta)+1)$ \cite{Richards1974}, meaning a diminishment of the coefficient $C_2$ and an enhancement of the coefficient $C_1$ in Equation~(\ref{eq:LW}). Indeed, such a crossover from the 2D to 3D type of the $\Delta H(\theta)$ dependence with lowering the temperature towards $T_{\rm N}$ was observed in the other family member of transition metal thiophosphates,  the vdW antiferromagnet \MPS \cite{senyk2023}. However, as discussed in Section~\ref{sec:X-band}, in \CCPS the 2D contribution to the angular dependence of the linewidth increases with lowering the temperature (Figure~\ref{fig:Adep_XBand}(d)). Such a different behavior can be related to the fact that in \MPS the predominant exchange interaction is antiferromagnetic whereas in \CCPS the much stronger intraplane exchange is ferromagnetic. Thus it boosts FM intraplane correlations towards $T_{\rm N}$ resulting in the long-range FM order of individual $ab$-planes which are AFM ordered along the $c^*$ direction. In this regard, considering both the anisotropic shift of the ESR line and the specific angular dependence of its width, \CCPS demonstrates in the bulk form  an intrinsically  2D FM spin dynamics very similar to the quasi-2D ferromagnet \CGT \cite{zeisner2019}. 
	
\subsection{Spin excitations and anisotropy in the AFM ground state}
\label{sec:AFM_ground_state}

The LSWT modeling of the AFMR  branches at the lowest measured temperature, as depicted in Figure \ref{fig:FDep_3K} with parameters summarized in Table~\ref{tab:parameters}, strongly suggests that \CCPS behaves as an easy-plane antiferromagnet  with a relatively small in-plane anisotropy where the $b$-axis is the easy-axis, conventionally termed as a biaxial two-sublattice antiferromagnet. The obtained $A_\mathrm{ex}$ sets the antiferromagnetic exchange strength between the neighboring layers in the \CCPS structure belonging to the opposite sublattices. $K_\mathrm{uniax}$ imposes an in-plane orientation of the sublattice magnetization vectors and $K_\mathrm{biax}$ yields their preferential orientation parallel to the $b$-axis. As can be seen above, this finding is strongly supported by the DFT calculations, yielding an antiferroelectric configuration as the lowest energy one. The calculated exchange and anisotropy parameters using this structural configuration, as well as the determined easy axis, match very well the results of the analysis of the experimental data (Table~\ref{tab:parameters}). 

\begin{figure*}[t]
	\centering
	\includegraphics[width=\linewidth]{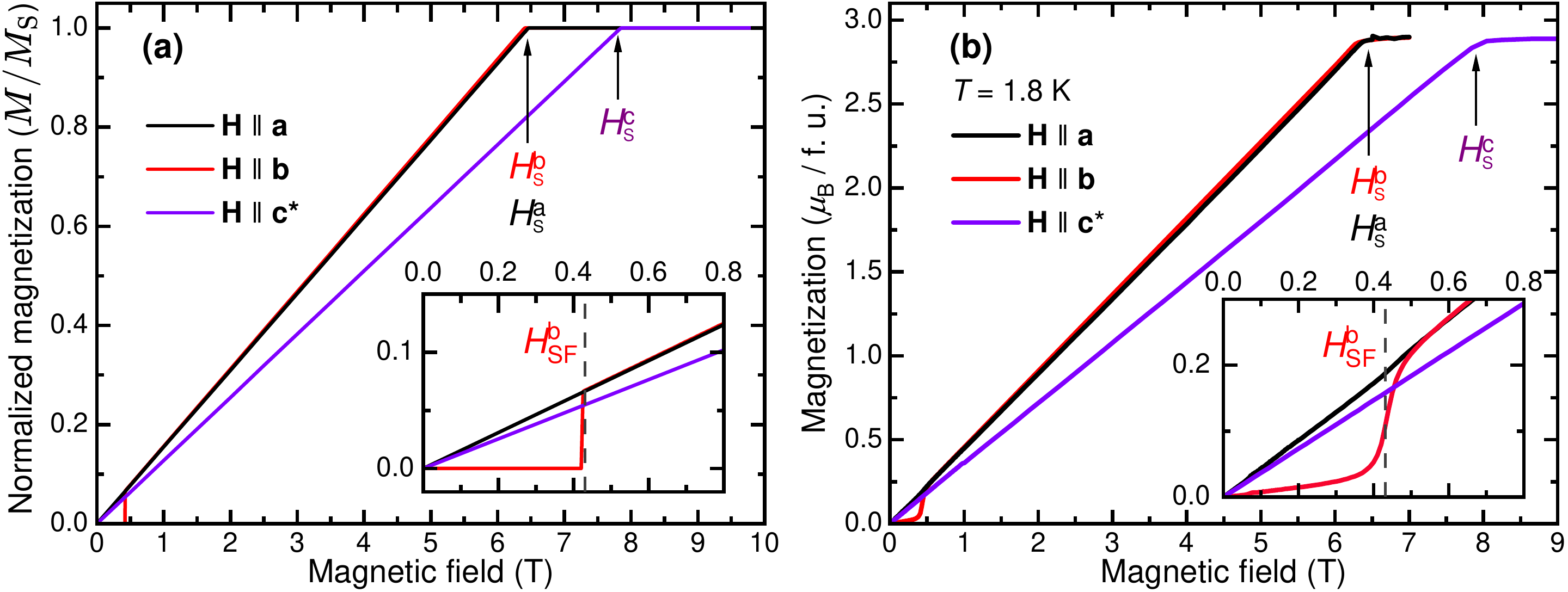}
	\caption{(a) Static magnetization normalized to its saturation value as a function of magnetic field at $T = 0$ \textit{calculated} by minimization of Equation~(\ref{Hamil}) with the parameters obtained from the modeling of the AFMR branches for $\bf H$ parallel to the $a$-, $b$- and $c^\ast$-axes, respectively. Arrows indicate the saturation fields $H\mathrm{_S^{b}}=6.4$\,T, $H\mathrm{_S^{a}}=6.5$\,T and $H\mathrm{_S^{c^*}}=7.8$\,T, and the spin-flop field $H\mathrm{_{SF}^{b}}=0.43$\,T. (b) Magnetization \textit{measured} at $T = 1.8$\,K with magnetic field $\bf H$ parallel to the $a$-, $b$- and $c^\ast$-axes, respectively.  } 
	\label{fig:Mag}
\end{figure*}

Specifically, both the HF-ESR and the DFT results reveal that \CCPS exhibits the AFMR characteristics typical for an A-type antiferromagnet, with a hard axis aligned along the $c^\ast$-axis determined by the positive sign of $K_{\rm uni}>0$ and an easy-axis aligned along the $b$-axis determined by the non-zero $K_{\rm biax}$ with $|K_{\rm biax}|\ll |K_{\rm uni}|$. The LSWT modeled resonance branch L$_{\rm c^\ast}$, as well as the experiments performed in the frequency domain (see Appendix~\ref{sec:FreqDom}), uncover the excitation gap $\Delta_{2} (H = 0) = 84$\,GHz  in the zero-field limit (Figure~\ref{fig:FDep_3K}(a)). When the external field is applied along the hard $c^\ast$-axis, $\Delta_{2}$  represents the minimum energy required to excite the ``in-phase'' precession of the sublattice magnetization vectors $\boldsymbol{M_1}$ and $\boldsymbol{M_2}$. Turning the field vector  to the in-plane easy $b$-axis significantly reduces the excitation energy of such an ``in-phase'' precession mode since in this configuration the magnetization vectors can be easily flopped from the antiparallel alignment already in small magnetic field (spin-flop field $\mu_{0}H_{\rm SF}\sim 0.4$\,T) (Figure~\ref{fig:Mag} and Refs.~\cite{Selter2023,wang2023}). The joint precession of the sublattice magnetization vectors with further increasing the field strength gives rise to the resonance branch L$_{\rm b}$ (Figure~\ref{fig:FDep_3K}(a)). Such time evolution of the dynamics of the sublattice magnetization vectors is presented as video files in Supporting Information.

The AFMR data for $\textbf{H}$ parallel to the easy axis \cite{axes_avsb} in the low-field part of the $\nu(H_{\rm res})$ diagram around $H_{\rm SF}$ were reported by Wang {\it et al.}~\cite{wang2023}. This result is depicted by open diamond symbols in Figure~\ref{fig:FDep_3K}(b). The modeled $\nu(H_{\rm res})$ dependence of the low-frequency AFMR signal based on the above analysis of the HF-ESR data reproduces this low-field AFMR data very well. The calculated branch L$_{\rm b}$ softens by reducing the field towards $H_{\rm SF}$ and the new branch L$^\prime_{\rm b}$ corresponding to the precession of the magnetization vectors in the spin-collinear phase emerges at $H \leq H_{\rm SF}$ ascending in energy and approaching the  excitation gap $\Delta_{1} (H = 0) = 13$\,GHz  in the zero-field limit.

At this point a substantial difference in the characteristics of the AFMR excitations of an easy-axis and an easy-plane antiferromagnet should be noted. While in a uniaxial easy-axis antiferromagnet the energies of the modes of the two sublattices in zero magnetic field are degenerate, the presence of biaxial anisotropy lifts this degeneracy yielding two zero-field excitation gaps as, for instance, observed in \MPS \cite{abraham2023}. However, in the case of a uniaxial easy-plane antiferromagnet, characterized by one gapped and one gapless spin-wave mode \cite{Turov}, the occurrence of the biaxial anisotropy lifts up in energy the gapless mode  introducing a second gap. In the case of \CCPS it amounts to $\Delta_{1} = 13$\,GHz, which is significantly smaller than the ``upper'' gap $\Delta_{2} = 84$\,GHz. Such a large difference in the gap values arises due to a large ratio $K_{\rm uni}/K_{\rm biax} \sim 50$, since the magnitude of an AFM excitation gap is approximately given by the geometric mean of exchange and anisotropy energies $\sqrt{AK}$.  

A significant difference in the energies of the two gapped AFM modes, and a different character of the respective excitations, make \CCPS an interesting candidate material for the use in prototype antiferromagnetic spintronic devices as a medium for transmitting the magnonic spin current without application of an external magnetic field. Generally, spin current, i.e., the flow of the spin angular momentum, can be generated by coherent excitation of monochromatic AFM magnons by microwaves or by thermal excitation. In the uniaxial easy-axis antiferromagnet the energies of the two magnon modes are degenerate at $H = 0$. The two modes correspond to the left-hand and right-hand circular precession of the sublattice magnetizations and carry opposite angular momenta. Excitation of these modes will thus result in the zero net spin current.
A finite field $H$ is required to lift the zero-field degeneracy of the spin-wave modes to enable a unidirectional flow of the spin current \cite{Rezende2016,Gomonay2018,Rezende2019,Han2023}. In an easy-plane antiferromagnet, such as \CCPS, the two modes at $H = 0$ are distinct in energy and consequently have different thermal populations. Their precession is nearly linear and carry vanishing angular momentum (see video files in Supporting Information). However, it has been shown that propagation of these two modes having different wavelengths  may yield a long-distance spin transport arising from their mixing and interference, in analogy to the birefringence effect in optics \cite{Han2020}. Indeed,
the modeling of the spin wave dispersion in different directions of the Brillouin zone of \CCPS presented in Appendix~\ref{sec:spinwaves} corroborates the results of the above discussion  and substantiates the conclusion on the strongly anisotropic character of the propagation of magnetic excitations in \CCPS.

With the parameters of Hamiltonian~(\ref{Hamil}) obtained in the modeling of the AFMR data the zero temperature magnetic field dependence of the magnetization $M(H)$ can also be calculated. That enables the determination of the saturation fields $H\mathrm{_S^{b}}=6.4$\,T, $H\mathrm{_S^{a}}=6.5$\,T and $H\mathrm{_S^{c^*}}=7.8$\,T for the three crystallographic directions and the spin-flop field for \HIIb $H\mathrm{^{b}_{SF}}=0.43$\,T, as shown in Figure~\ref{fig:Mag}(a). The obtained values are in a very good agreement with the experimental $M(H)$ data measured at finite temperatures below $T_{\rm N}$ (Figure~\ref{fig:Mag}(b), Appendix~\ref{sec:abc_prop} and Refs.~\cite{park2022,Selter2023,wang2023}) thereby additionally validating the results of the analysis of the AFM resonance branches. Moreover, our modeling reproduces well the characteristically shaped angular dependence of the magnetization measured at different magnetic fields below, above and close to the spin flop transition (Figure~\ref{fig:appendix_magn}(c) in Appendix~\ref{sec:appendix_magn}). The anisotropy of the saturation field is due to the interplay of the exchange interaction, the applied field and the magnetic anisotropies $K_\mathrm{uniax}$ and $K_\mathrm{biax}$, yielding the hierarchy $H\mathrm{_S^{b}} < H\mathrm{_S^{a}} < H\mathrm{_S^{c^*}}$. In the \HIIb configuration, both anisotropies aid the field to polarize the spin moments whereas for \HIIa configuration, only the first term works in favor of the field. In contrast for \HIIc configuration, both terms work against the field, hence increasing $H_\mathrm{S}$. It has been argued that the curvature in the isothermal magnetization at a low field corresponding to $H_\mathrm{SF}$ (inset in Figure~\ref{fig:Mag}(b)) is not the conventional spin-flop transition  but a feature of magnetoelectric coupling \cite{lai2019}. However, this spin-flop transition can be reproduced with the canonical approach presented here (Section~\ref{sec:FDep_3K}), assuming the presence of an in-plane magnetocrystalline anisotropy ($K_{\rm biax} \neq 0$ in Equation~(\ref{Hamil})) that gives rise to both the excitation gap $\Delta_{1}$  and the spin-flop at $H\mathrm{^{b}_{SF}}$ without invoking more complex scenarios. Importantly, the presence of the in-plane anisotropy is also suggested by our DFT calculations presented above.

Using the determined exchange energy density $A_\mathrm{ex}= 2.3 \times 10^6$ \,erg/cm$^3$ in Hamiltonian~(\ref{Hamil}) which defines the AFM interaction energy between the layers of \CCPS one can make a mean-field theory estimate of the Curie-Weiss constant given by $\Theta_{\rm CW} = A_\mathrm{ex}\times(C/M_0^2)$, where $C$ represents the Curie constant. This provides an estimation of the average energy scale for the interlayer AFM exchange interactions for \CCPS corresponding to $\Theta_{\rm CW}^{\rm inter} \sim - 7$\,K. Notably, the average Curie-Weiss constant containing both inter- and intraplane interactions  obtained from the temperature dependence of the inverse static susceptibility $\chi^{-1}(T) = (T-\Theta_{\rm CW}^{\rm aver})/C$ is positive, i.e. FM, and is much larger in magnitude amounting to $\Theta_{\rm CW}^{\rm aver} \sim 30$\,K (see Figure~\ref{fig:appendix_magn}(b) in Appendix~\ref{sec:appendix_magn} and \cite{Colombet1982,Kleemann2011,Selter2023}). This is expected for an A-type AFM order of the spin lattice, as in the case of \CCPS where, the spins are strongly FM-coupled in the planes and only weakly AFM interact between the planes. If to assume  for simplicity that every spin in a selected plane interacts antiferromagnetically with only two spins (one in the upper plane and the other one in the plane beneath) with the Heisenberg exchange $\mathcal{H} = J_{\rm inter} \sum S_iS_j$, the mean-field estimate yields $J = 3|\Theta_{\rm CW}^{\rm inter}|/zS(S+1) \approx 2.8\,{\rm K} = 0.24$\,meV in a reasonable agreement with the DFT calculation of the interlayer exchange in Ref.~\cite{park2022}. Note that, here $z=2$ is the number of neighbors and $S = 3/2$ is the Cr$^{3+}$ spin value.

A remarkable observation in our HF-ESR study of \CCPS  at $T\ll T_{\rm N}$ is a  transformation of the  AFM branches L$_{\rm b}$ and L$_{\rm c^\ast}$ into the FM branches in a field range $\sim 3.5 - 7$\,T  (Figure~\ref{fig:FDep_3K}(a)). With increasing the field strength both L$_{\rm b}$ and L$_{\rm c^\ast}$ intersect each other and evolve parallel to the paramagnetic branch L$_{\rm par}$. This is typically not observed in conventional antiferromagnets where, in accordance with the classical textbook analytical treatments \cite{Turov} the solutions are simplified assuming that the exchange term $A_{\rm ex}$ in Hamiltonian~(\ref{Hamil}) is by orders of magnitude larger than any magnetic anisotropy energy $K$. This holds for most of 3D antiferromagnets and also for some quasi-2D vdW antiferromagnets such as, e.g., \MPS, which belongs to the same family of transition metal thiophosphates and shows a conventional behavior of the AFM branches \cite{abraham2023}. In such cases, L$_{\rm b}$ and L$_{\rm c^\ast}$ asymptotically approach  L$_{\rm par}$ from above and below, respectively, but never cross. Our numerical modeling of the branches solves this paradox exactly reproducing this experimental observation and showing that $A_{\rm ex}$ is only about  five times larger than $K_{\rm uni}$. The closeness of these quantities  produces a rare scenario where the magnetic field can {\it qualitatively} change the type of magnetic excitations  from AFM to FM. The FM layers in \CCPS can be considered as effectively decoupled from each other by a moderate magnetic field and demonstrate spin dynamics indistinguishable from a genuine ferromagnet. Indeed, switching off the interlayer exchange, i.e., reducing the system to a one-sublattice ferromagnet by completely omitting the interlayer AFM exchange term in Hamiltonian~(\ref{Hamil}), exactly reproduces the high-field behavior of both branches (Figure~\ref{fig:FDep_3K}(a)) which in this regime corresponds to excitations of the easy-plane ferromagnet. Another example of such a transformation was observed in the HF-ESR experiments on the vdW magnet MnBi$_2$Te$_4$ \cite{alfonsov2021}.

\section{Conclusion}

In summary, we have studied the low-energy spin dynamics in the van der Waals antiferromagnet \CCPS featuring interpenetrating antiferroelectric Cu$^{1+}$ and antiferromagnetic Cr$^{3+}$ lattices with ESR spectroscopy at frequencies up to 330\,GHz, at magnetic fields up to 16\,T, and in the temperature range 3 -- 300\,K. Our major findings are: 
\begin{itemize}
	
\item[(i)] 
In the paramagnetic regime above the AFM ordering temperature $T_{\rm N} \approx 30$\,K the temperature, magnetic field and angular dependences of the ESR line shift and its width evidence inherently two-dimensional short-range FM spin correlations persistent in the planes of \CCPS at $T \gg T_{\rm N}$. However, the ESR data do not reveal  signatures of magneto-electric coupling in \CCPS within experimental error bars indicating its possible smallness usually expected for type-I multiferroics;
\item[(ii)] 
In the AFM ordered state below $T_{\rm N}$ mapping of the frequency versus magnetic field diagram of the resonance modes for different orientations of the magnetic field and its analysis in the frame of the linear spin wave theory enabled us to estimate several magnetic energy scales, such as the interlayer exchange constant, $A_\mathrm{ex}= 2.3 \times 10^6$ \,erg/cm$^3$ $= 2.3 \times 10^5$\,J/m$^3$, the uniaxial anisotropy constant, $K_\mathrm{uniax} = 5 \times 10^5$\,erg/cm$^3 = 5 \times 10^4$\,J/m$^3$ and the biaxial anisotropy constant, $K_\mathrm{biax} = 1 \times 10^4$\,erg/cm$^3= 1 \times 10^3$\,J/m$^3$, and to identify two magnon excitation modes in the zero-field limit with energies $\Delta_{1} = 13$\,GHz and $\Delta_{2} = 84$\,GHz. Importantly, these parameters are in a very good agreement with the results of our DFT calculations for the antiferroelectric structural configuration. An estimate of the Heisenberg exchange constant between the Cr spins in the neighboring planes yields the value $J \approx 2.8\,{\rm K} = 0.24$\,meV; 
\item[(iii)] 
The relative closeness of the energy scales of the interlayer exchange and the uniaxial magnetic anisotropy enabled us to observe a remarkable effect of the field tuning of the collective spin excitations from the AFM-type at small fields to the FM-type in strong fields;
\item[(iv)] 
The above observations highlight new interesting magnetic functionalities of \CCPS that may make it relevant for prototype magnonic devices. Specifically, the non-degeneracy of the magnon excitation modes in zero magnetic field offers a possibility to use \CCPS as a transmitter of the spin current, i.e., the flow of the spin angular momentum. This can be done by exciting the magnons with distinct energies $\Delta_{2}$ and $\Delta_{1} \ (\ll  \Delta_{2})$  without the application of an external magnetic field which in many cases is required to lift the modes' degeneracy \cite{Gomonay2018,Han2023}. Furthermore, the application of moderate magnetic fields enables a controllable switching of the character of the spin excitations between AFM and FM types of spin dynamics which can be treated as different ``channels'' for the transmission of the spin current in the same material. Indeed, the modeling of the spin wave dispersion reveals a strongly anisotropic character of the propagation of magnetic excitations in \CCPS and calls for a detailed study of the momentum-dependent spin dynamics in this material by inelastic neutron scattering.

\end{itemize}

\section{Experimental and computational details}
\label{sec:expt}

\

\subsection*{Samples}
High quality single crystals of \CCPS studied in this work were synthesized using the chemical vapor transport technique with iodine as the transport agent. All the detailed characterization, including growth, morphology, compositional, and structural data, is described in Ref.~\citenum{Selter2023}.

\subsection*{Crystal orientation via X-ray single crystal diffraction}
Several single crystals were selected for the ESR and magnetization measurements and oriented using the Bruker AXS Kappa Apex II diffractometer. For this purpose the crystals were mounted on a crystal holder, short data collection was performed at ambient temperature to determine the unit cell. The best single crystals gave about 80\,\% of reflections that fit the monoclinic C unit cell determined in Ref.~\cite{maisonneuve1993}, though the size of the crystals ($\sim 1.5$\,mm) was larger than the diameter of the X-ray beam. The orientation of the single crystals was determined using the “index crystal faces” tool incorporated into the APEX4 v2021.10-0 Software \cite{APEX4}. All single crystals had the same platelike habitus with the  $c^\ast$-axis being perpendicular to the plate. Due to the irregular shape of the plates, the directions of $a$ and $b$ axes were easy to align respective to the magnetic field applied in the ESR and magnetometry setups. As the result the crystallographic $b$-axis was identified as the magnetically easy direction. The crystal structure in Figure~\ref{fig:Structure} and Figure~\ref{fig:CuConfiguration} was drawn using the Vesta software \cite{Vesta}.

\subsection*{Static magnetometry}
Low-field and field-angle-dependent DC magnetic studies were conducted using the Quantum Design Magnetic Property Measurement System (MPMS3) with the horizontal rotator option. The temperature-dependent magnetic susceptibility was measured using the MPMS3, by employing  a quartz paddle for in-plane measurements and a quartz paddle with two quartz half-ciliynders holding the crystal vertically for out-of-plane measurements. High-field magnetization measurements for the \HIIc configuration were performed using the Quantum Design Physical Property Measurement System (PPMS) with a quartz paddle in a similar configuration.

\subsection*{X-Band electron spin resonance}
The low-frequency ESR measurements were performed using the Bruker X-band ESR-spectrometer operating at a constant frequency of $\nu$ = 9.56\,GHz and with magnetic fields up to 9\,kOe. The measurement temperature between 4\,K and 300\,K is reached using a $^4$He-gas flow cryostat from Oxford Instruments. 

In this setup the absorption derivative of the ESR spectra is measured. Therefore the fitting function reads:
\begin{align}
S_\mathrm{abs} =& \frac{-2x}{(1+x^2)^2} + \frac{-2y}{(1+y^2)^2}
\end{align}
where
\begin{align}
x = \frac{2(H+H_{res})}{\Delta H}, \hspace{5 pt}
y = \frac{2(H-H_{res})}{\Delta H}\ .
\label{eq:xy}
\end{align}

\subsection*{High-frequency electron spin resonance (HF-ESR)} 
The high frequency ESR measurements were performed using a home-made high-field/high-frequency electron spin resonance spectrometer. Here, a vector network analyzer (PNA-X) from Keysight Technologies with the extensions from Virginia Diodes, Inc. (VDI) were used for the generation and detection of microwaves (MW) in the frequency range from 75\,GHz to 330\,GHz. For continuous magnetic field sweeps, a superconducting magnet from Oxford Instruments was used. The homemade probehead with oversized waveguides was inserted into a variable temperature insert (VTI) of a He$^{4}$ cryostat (from Oxford Instruments) to obtain stable temperatures of the sample from 300\,K down to 3\,K. All measurements were performed by continuously varying the magnetic fields from 0\,T to 16\,T and back at a constant temperature and frequency.

The ESR signal in this case can be described by the following fitting function:
\begin{align}
S_\mathrm{D}^\mathrm{amp}=\frac{2A_0}{\pi}(S_\mathrm{abs}\sin\alpha+S_\mathrm{disp}\cos\alpha)+S_o+S_s\cdot H 
\label{eq:lorentzian}
\end{align}
where $A_0$ is the amplitude of the signal, $S_0$ is the offset and $S_s$ is the linear background in the detected signal. Here $S_\mathrm{abs}$ and $S_\mathrm{disp}$ are related by the Kramers-Kronig relation and can be expressed as:
\begin{align}
S_\mathrm{abs} =& \frac{1}{1+x^2} + \frac{1}{1+y^2}\\
S_\mathrm{disp} =& \frac{x}{1+x^2} + \frac{y}{1+y^2}\ ,
\end{align}
where $x$ and $y$ are defined by Equation~(\ref{eq:xy}). All the spectra measured in this work were fitted with this function to obtain the parameters $H_{res}$ and $\Delta H$.

\subsection*{LSWT}
Numerical energy minimization and calculations of the resonance modes using linear spin wave theory was performed using Julia language \cite{Bezanson2017}. The resulting dynamics of the sublattice magnetization vectors was animated using the package Makie.jl \cite{Danisch2021}.

\subsection*{DFT}
The DFT calculations were performed using the full-potential local-orbital  code FPLO version 18.00-57 \cite{koepernik1999full}. The generalized gradient approximation (GGA)~\cite{PBE96} is adopted for the exchange and correlation functional, and the GGA+$U$ method was used to account for the strong correlations of $3d$ electrons of Cr ($U$\,=\,4.5 eV, $J$\,=\,1 eV, the fully localized limit for the double-counting correction). Optimisation of $U$ used in DFT+$U$ calculations is presented in Appendix~\ref{Sec:DFT_appendix}. To analyze the total energies of configurations with different Cu displacements, we performed scalar relativistic calculations on the mesh of 4$\times$4$\times$2 $k$-points. For the configuration with the lowest energy, we performed a full structural optimization using the SCAN+$U$ functional \cite{SCAN} implemented in VASP version 6.4.2 \cite{VASP, VASP_2} employing projector-augmented wave potentials~\cite{VASP_PAW}. The resulting crystal structure is provided in Table \ref{tab:scan-lattice2} in Appendix~\ref{Sec:DFT_appendix}. For this structure,  we calculated the effective interlayer magnetic exchange, by comparing the GGA+$U$ total energies of $A$-type antiferromagnetic and ferromagnetic configurations; these calculations were performed on a mesh of 4 $\times$ 6 $\times$ 10 $k$-points. To calculate the magnetic anisotropy energies, we used full-relativistic magnetic GGA+$U$ calculations on a $k$-mesh of 8 $\times$ 4 $\times$ 4 points with stringent convergence criteria (10$^{-8}$ Hartree for total energy and 10$^{-6}$ for charge density).

\medskip
\textbf{Acknowledgments} \par

The authors express gratitude to Miroslav Po\v{z}ek for very fruitful scientific discussions and  to Thomas Doert for the permission to use the Bruker AXS diffractometer. This work was supported by the Deutsche Forschungsgemeinschaft (DFG, German Research Foundation) through grants No. 447482487 (KA 1694/12-1), 499461434 (AL 1771/8-1), 455319354, SFB1143 (project. No. 247310070), the Dresden-Würzburg Cluster of Excellence (EXC 2147) ``ct.qmat - Complexity and Topology in Quantum Matter'' (project-id 390858490) and by the German Federal Ministry of Education and Research (BMBF) through project GU-QuMat (01DK240008). The work in Zagreb was supported by the Croatian Science Foundation through grant No. HRZZ-UIP-2020-02-9494.

\pagebreak

\section{Appendix}
\label{Sec:appendix}

\subsection{Static magnetometry}
\label{sec:appendix_magn}

\begin{figure*}[!t]
	\centering
	\includegraphics[width=0.95\linewidth]{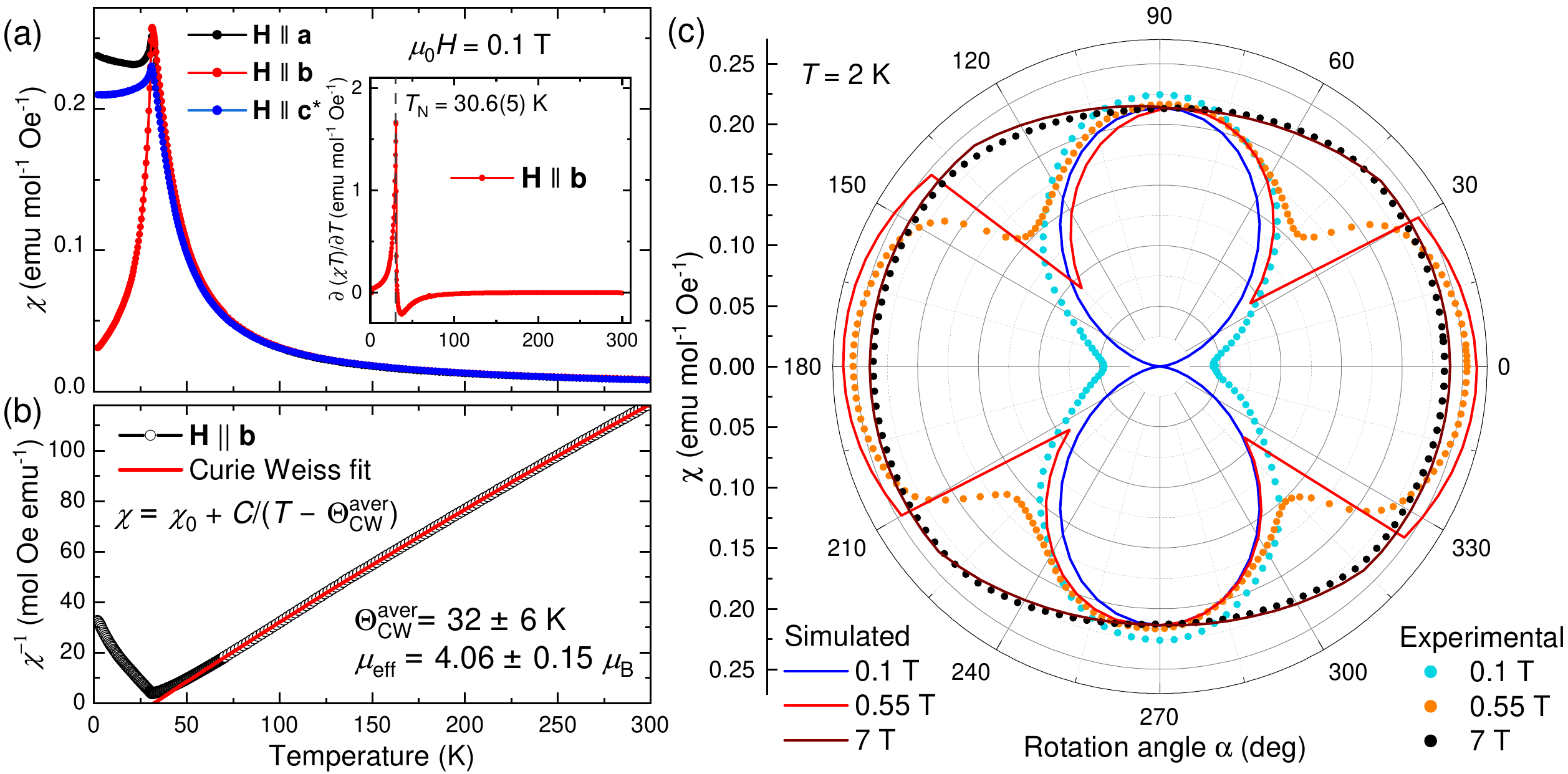}
	\caption{Magnetic susceptibility $\chi (T)$ (a) and $1/ \chi (T)$ (b) as a function of temperature measured at $\mu_{0}H = 0.1$\,T. Inset in (a) shows the derivative $\partial \chi / \partial T$ with a sharp discontinuity at $T_{\rm N}$. (c) Angular dependence of the magnetization in the $bc$-plane measured at three magnetic field values, below the spin flop at $\mu_{0}H = 0.1$\,T, just above the spin flop at $\mu_{0}H = 0.55$\,T, and far above the spin flop at $\mu_{0}H = 7$\,T. The dotted lines show the experimental data, while the solid lines correspond to the simulated data (see text for details).}
	\label{fig:appendix_magn}
\end{figure*}

Figure~\ref{fig:appendix_magn}(a) shows the magnetic susceptibility $\chi (T)$ measured under a magnetic field of $0.1$\,T applied along three crystallographic directions: $a$, $b$, and $c^{*}$. The sharp decrease in $\chi (T)$ below $T_{\rm N}$ for the configuration \HIIb suggests the alignment of antiferromagnetically coupled spins, indicating that the magnetic easy axis lies along the crystallographic $b$-axis. The AFM ordering with a transition temperature $T_{\rm N} = 30.6 \pm 0.5$\,K, in turn, is characterized by a cusp, as shown in the inset.

As can be seen in Figure~\ref{fig:appendix_magn}(b), showing the $\chi^{-1} (T)$ dependence, above the transition temperature $T_{\rm N}$, the magnetic susceptibility follows the Curie-Weiss law, $\chi = \chi_0 + C/(T-\Theta^{\rm aver}_{\rm CW})$. Here $\chi_0$ is the temperature-independent term, $C$ is the Curie-Weiss constant, and $\Theta^{\rm aver}_{\rm CW}$ the Curie-Weiss temperature. $\chi^{-1} (T)$ exhibits a linear dependence down to approximately 60 K, below which short-range correlations cause deviations from the Curie-Weiss law. The Curie-Weiss fit yields $\chi_0 = (7.4 \pm 0.2) \times 10^{-4}$\,emu/(mol Oe), the effective magnetic moment $\mu_{\rm eff} = 4.06 \pm 0.15$\,$\mu$B, and $\Theta^{\rm aver}_{\rm CW} = 32 \pm 6$\,K. The effective magnetic moment value is close to the expected value for Cr$^{3+}$ ions with $S=3/2$ and $g=2$. The positive Curie-Weiss temperature indicates the predominance of ferromagnetic interactions, consistent with the A-type antiferromagnetic order (ferromagnetic layers stacked antiferromagnetically) in \CCPS.

Furthermore, as shown in Figure~\ref{fig:appendix_magn}(c), the angular dependence of the magnetic susceptibility in the $bc$-plane at $T = 2$\,K exhibits two-fold symmetry below the spin-flop field ($\mu_{0}H = 0.1$\,T), four-fold symmetry just above the spin-flop field ($\mu_{0}H = 0.55$\,T), and elliptic symmetry at $\mu_{0}H = 7$\,T, marking the anisotropy present between the $b$- and $c^{\ast}$-axes. The angular dependence of magnetic susceptibility was simulated using the parameters obtained with the LSWT numerical modeling of the ESR data, reproducing the symmetries observed in the experimental data with excellent agreement.

\subsection{Frequency domain ESR measurements}
\label{sec:FreqDom}

\begin{figure*}[!t]
	\centering
	\includegraphics[width=1.0\linewidth]{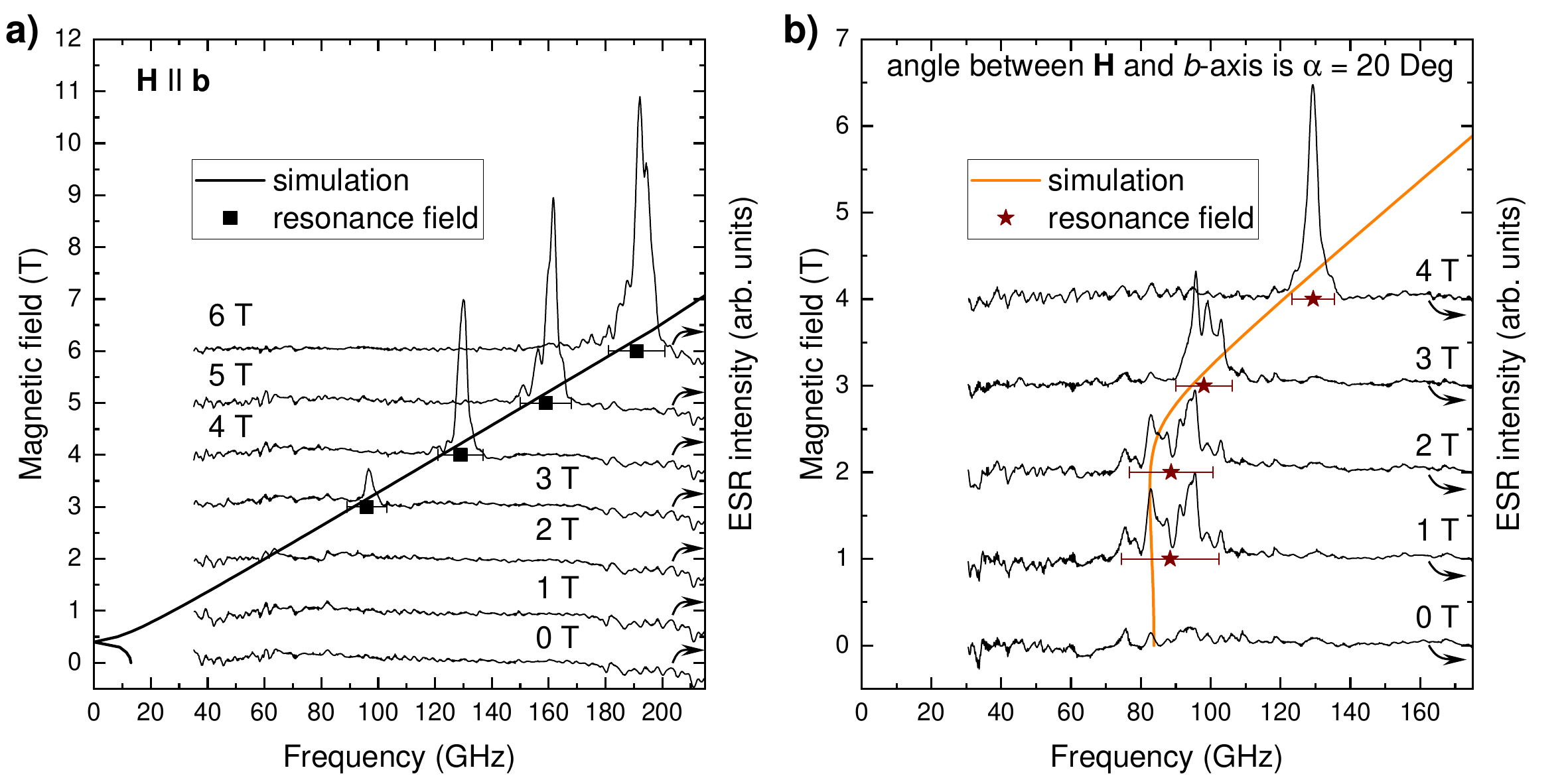}
	\caption{Frequency domain measurements performed at a temperature of 1.5\,K. a) \HIIb configuration of magnetic field. b) Magnetic field was set to make an angle of  20 degrees with the $b$-axis in the $bc^*$-plane.}
	\label{fig:Freq_Sweep}
\end{figure*}

Frequency domain ESR measurements \cite{Rogic2024}, i.e., measurements where the microwave frequency is swept whereas the magnetic field is kept constant, were performed in order to directly measure the zero-field energy gap for spin wave excitations in the ordered state of \CCPS. The attempt to perform such measurements in the configuration \HIIc was not successful, possibly due to inefficient coupling between microwaves and spin wave excitations, which are linear in the zero-field limit (animations of the dynamics see in Supporting Information). The AFMR signal could be detected at finite fields in the \HIIb geometry, with its amplitude drastically decreasing when approaching zero magnetic field (Figure~\ref{fig:Freq_Sweep}(a)). However, the successful measurement of the spin wave gap at the smallest field was performed in the configuration when the magnetic field was set to make an angle of 20 degrees with the $b$-axis in the $bc^*$-plane, as can be seen in Figure~\ref{fig:Freq_Sweep}(b).

\FloatBarrier

\subsection{Shape anisotropy}
\label{Sec:Shape_anisotropy}
\begin{figure}[!t]
	\centering
	\includegraphics[width=1.0\linewidth]{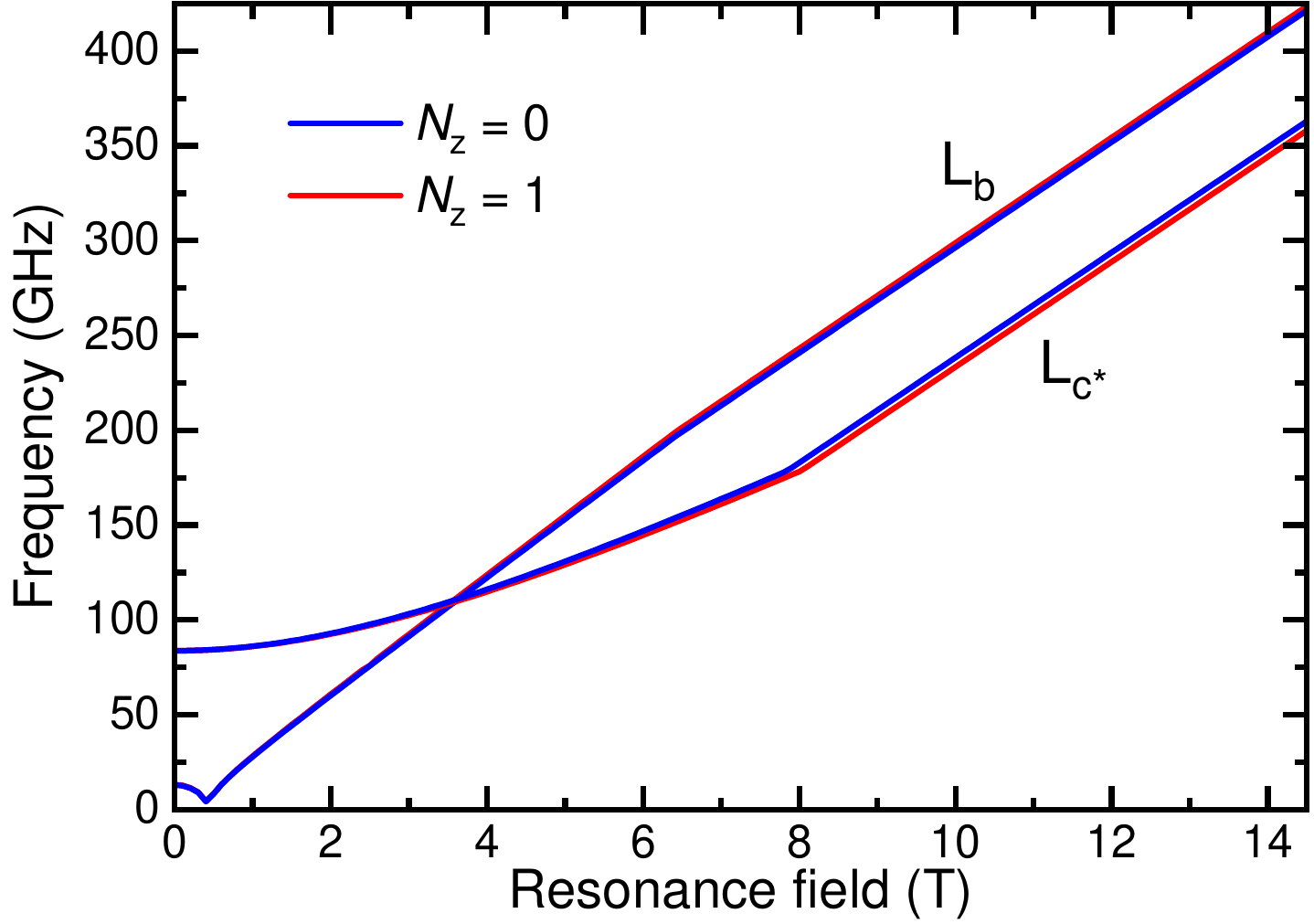}
	\caption{Modeled $\nu(H_{\rm res})$ dependence of the AFMR branches for \CCPS for two limiting values of the shape factor $N_{\rm z}$ of 0 and 1, blue and red lines, respectively.} 
	\label{fig:ShapeAnisotropy}
\end{figure}
The influence of the shape anisotropy due to the demagnetization fields for a specific shape of the single crystal of \CCPS was evaluated with the LSWT numerical modeling by adding to Equation~(\ref{Hamil}) the shape anisotropy term $2\pi N_{\rm z}(M_{\rm 1z} + M_{\rm 2z})^2$. This effect was calculated in the experimental range of the applied field and frequency for two limiting values of the demagnetization factor $N_\mathrm{z} = 0$ (sphere) and $N_\mathrm{z}=1$ (infinite thin plate). The observed effect was found to be minimal as is illustrated in Figure~\ref{fig:ShapeAnisotropy}. Hence, any influence of the shape anisotropy is neglected for this study.

\FloatBarrier

\subsection{DFT calculations}
\label{Sec:DFT_appendix}

\begin{figure}[!h]
	\centering
	\includegraphics[width=0.5\textwidth]{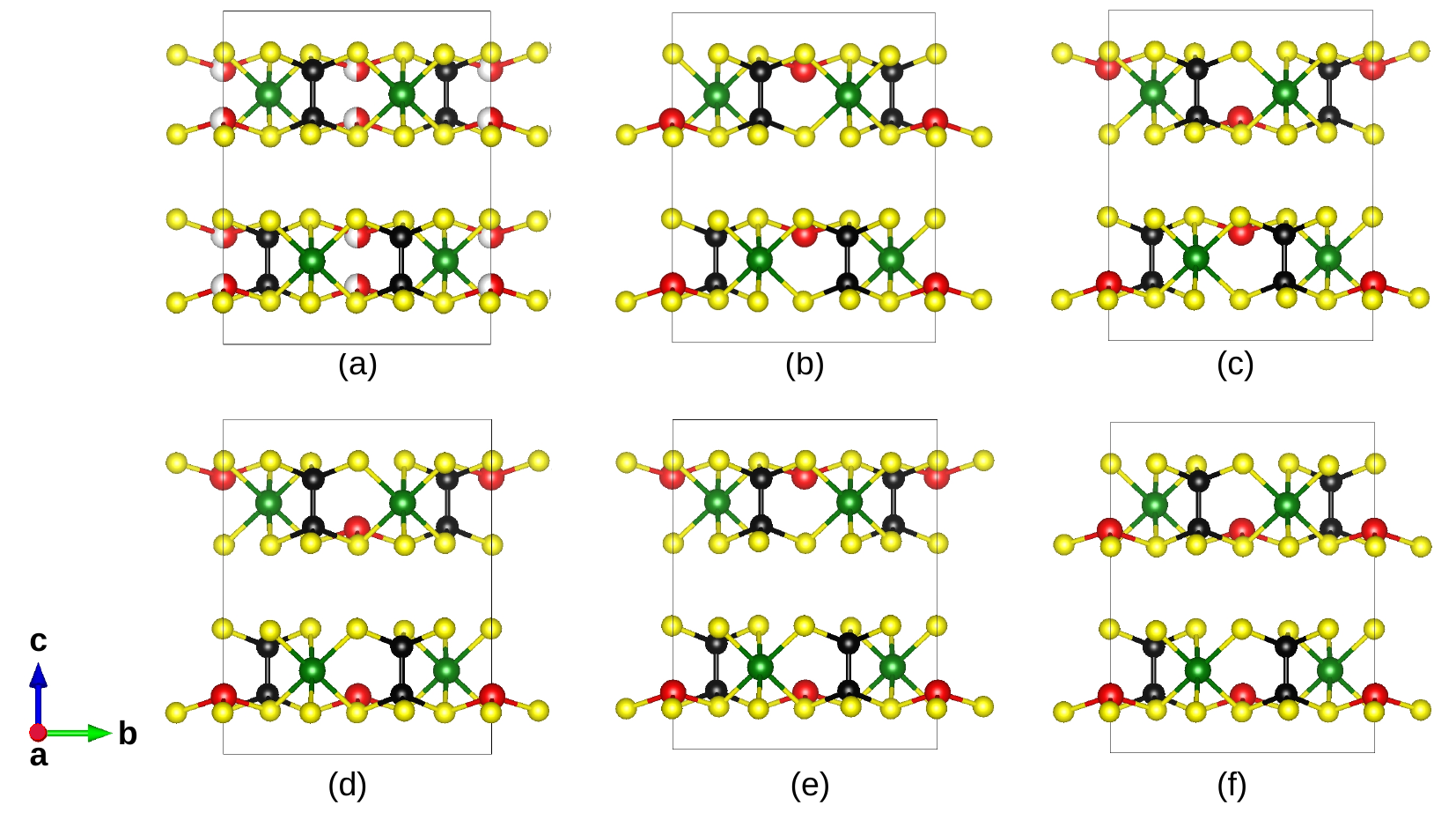}
	\caption{Crystal structures of CuCrP$_2$S$_6$ with different Cu configurations. Cu atoms are shown in red and plotted together with other atoms -- S (Light yellow), P (black), Cr (green). (a) Each Cu site has 50\% occupancy. (b) Cu atoms are located at the top and the bottom of S$_6$ octahedra alternatively, making both layers antiferroelectric. Cu-Cu chains in both layers are in phase in the configuration. While the configuration (c) is similar to (b), the difference is that Cu-Cu chains are out of phase in this configuration. (d) Cu atoms are located at the top and bottom of S$_6$ octahedra alternatively in the upper layer, while all Cu atoms are located at the bottom of S$_6$ octahedra in the lower layer, leading to a ferrielectic configuration. (e) All Cu atoms are located at the top of the S$_6$ octahedra in the upper layer, while all Cu atoms are located at the bottom of the S$_6$ octahedra in the lower layer, resulting in a overall antiferroelectric configuration. (f) Both layers are ferroelectric, giving rise to ferroelectric configuration. }
	\label{fig:CuConfiguration}
\end{figure}

\begin{table*}[!t]
	\centering
	\begin{tabular}{r c c c | r c c c}
		atom & $x/a$ & $y/b$ & $z/c$ & atom & $x/a$ & $y/b$ & $z/c$  \\
		\hline
		Cu1 &  0.89600743&  0.75106954&  0.21627343& 	S3 &  0.87397991&  0.93753709&  0.38050696  \\
		Cu2 &  0.60399257&  0.24893046&  0.28372657& S4 &  0.62602009&  0.06246291&  0.11949304 \\
		Cr1 &  0.00096754&  0.58274214&  0.75103359& S5 &  0.37845530&  0.42296160&  0.39378018 \\ 
		Cr2 &  0.49903246&  0.41725786&  0.74896641& S6 &  0.12154470&  0.57703840&  0.10621982  \\
		P1 &  0.41568651&  0.58038756&  0.22817050& S7 &  0.37995721&  0.93088438&  0.89712581  \\
		P2 &  0.08431349&  0.41961244&  0.27182950&  S8 &  0.12004279&  0.06911562&  0.60287419  \\
		P3 &  0.42190506&  0.08712098&  0.71357843& S9 &  0.88461306&  0.24840074&  0.36037686  \\
		P4 &  0.07809494&  0.91287902&  0.78642157& S10 &  0.61538694&  0.75159926&  0.13962314  \\
		S1 &  0.88283616&  0.42752819&  0.87626861& S11 &  0.87390215&  0.74124627&  0.84212588  \\
		S2 &  0.61716384&  0.57247181&  0.62373139& S12 &  0.62609785&  0.25875373&  0.65787412  \\
	\end{tabular}
	\caption{Atomic coordinates in the SCAN+$U$ optimized structure, space group $P2_1(4)$, $a$ = 13.424886\,\r{A}, $b$ = 10.301879\,\r{A},  $c$ = 5.940316\,\r{A}, $\beta$ = 98.837227$^{\circ}$.}
	\label{tab:scan-lattice2}
\end{table*}

\subsubsection*{Estimation of $U$ used in DFT+$U$ calculations} 

Figure~\ref{fig:CuConfiguration} shows crystal structures considered in the DFT calculations.
Total energies calculated in DFT+$U$ naturally depend on the $U$ value, hence all quantities computed using such energies -- magnetic exchange couplings and anisotropy parameters -- also have this dependence. To estimate the optimal $U$, we used the value of the Weiss temperature $\theta \simeq$35 K obtained from a Curie-Weiss fit to the magnetic susceptibility (Appendix~\ref{sec:appendix_magn}). This value is related to the sum of all exchange couplings $J_i$ (in K) as:

\begin{equation}
\theta_{\text{W}} = -\frac{S(S+1)}{3}\sum_{i}z_iJ_i, \end{equation}

where $z_i$ is the multiplicity. Since the in-plane exchanges are dominant, the interplane exchanges were neglected and the Weiss temperature from two inequivalent ferromagnetic nearest-neighbor in-plane exchanges $J_1$ and $J_1'$ was estimated as

\begin{equation}
\theta_{\text{W}} = -\frac{5}{4}(4J_1 + 2J_1'), \end{equation}

where we used $S=\frac32$. In this way, the best agreement for $U$ = 1.75 eV was obtained (Table~\ref{tab:U_dep}).

\begin{table*}[!t]
	\centering
	\caption{Magnetic exchange and Curie--Weiss temperatures as a function of Hubbard \( U \).} \begin{tabular}{c|ccccccccc} \hline U value (K) & 1 & 1.5 & 2 & 2.5 & 3 & 3.5 & 4 & 4.5 & 5 \\ \hline
		
		$J_1$ & -2.12 & -4.20 & -5.82 & -7.13 & -8.20 & -9.11 & -9.89 & -10.56 & -11.14 \\
		$J_1'$ & -1.62 & -3.15 & -4.36 & -5.36 & -6.19 & -6.91 & -7.56 & -8.14 & -8.66 \\
		
		$\theta_{\text{W}}$ & 14.65 & 28.89 & 40.03 & 49.04 & 56.47 & 62.82 &
		68.34 & 73.15 & 77.37 \\
		\hline
	\end{tabular}
	\label{tab:U_dep}
\end{table*}

\FloatBarrier

\subsection{Angular dependence of $\mathbf{H_{res}}$ in the $\mathbf{bc^*}$-plane}
\label{sec:Hres_angular}

The two-fold symmetry of the $H_{\rm res}(\alpha)$ dependence is expected to arise due to the temperature dependent uniaxial anisotropic internal field, or single-ion anisotropy, and the temperature independent anisotropic $g$-factor. To analyze the deviation of the $H_{\rm res}(\alpha)$ from the two-fold symmetry, the dependences shown in Figure~\ref{fig:Adep_XBand}(c) were fitted with the phenomenological function comprising the two-fold and four-fold harmonics as:
\begin{align}
	\label{eq:Hres}
	\mu_0 H_{\rm res}(\alpha) = A(T)\cos(2\alpha) + B(T) \cos(4\alpha) + \mu_0 H_{\rm res}^{\rm mean}(T)
\end{align}
Here, $A$ and $B$ are the weighting coefficients of the respective harmonics and $H_{\rm res}^{\rm mean}$ is the mean resonance field at a given temperature.  
\begin{figure}[!h]
	\centering
	\includegraphics[width=1.0\linewidth]{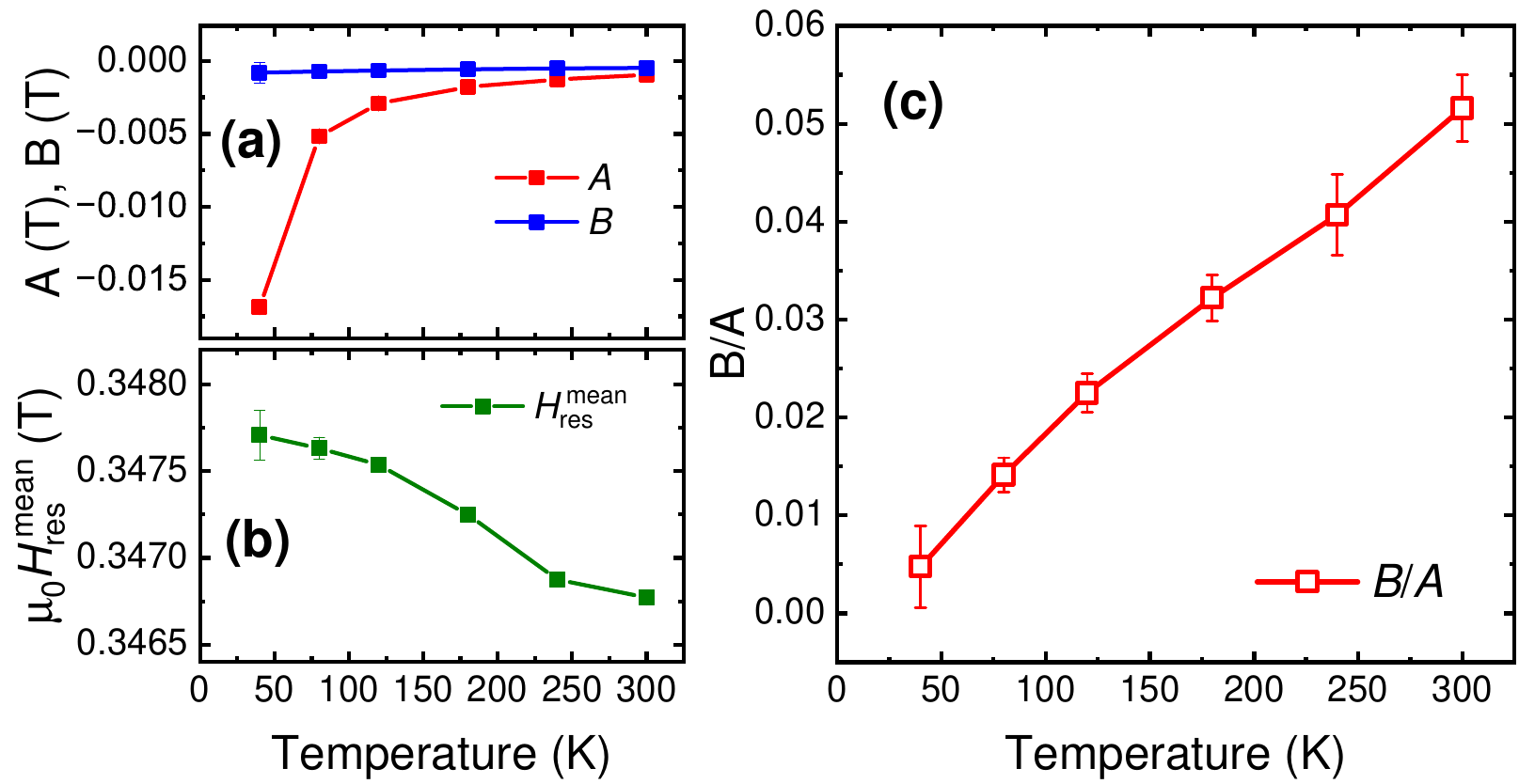}
	\caption{Temperature dependence of coefficients in Equation~(\ref{eq:Hres}):  $A$ and  $B$ (a), $H_{\rm res}^{\rm mean}$ (b) , and of the relative contribution of $B$ with respect to $A$ (c). Solid lines are guides to the eye.}
	\label{fig:Adep_Hres}
\end{figure}
Their temperature dependence is plotted in Figure~\ref{fig:Adep_Hres}(a,b). While the coefficient $A$ as well as $H_{\rm res}^{\rm mean}$  strongly decrease with increasing temperature due to a ceasing of  2D FM spin correlations which give rise to the persisting anisotropic internal field above $T_{\rm N}$, coefficient $B$ appears to be only weakly dependent on temperature. Thus, the relative weight of the four-fold symmetry term $B/A$ in the $H_{\rm res}(\alpha)$ dependence gradually increases with temperature, as shown in Figure~\ref{fig:Adep_Hres}(c), giving rise to a local minimum of this dependence at $\alpha = 90^\circ$ in the high-temperature limit (Figure~\ref{fig:Adep_XBand}(c)). The origin of this weakly temperature dependent four-fold harmonic, which cannot be related to any local crystal field effect, is yet to be clarified, since for  individual Cr$^{3+}$ ions with $S =3/2$, harmonics only up to third order may have nonzero values \cite{Abragam2012}. 
The smooth linear relative increase of $B$ with respect to $A$ also  rules out the possibility that the change of the symmetry of the $H_{\rm res}(\alpha)$ dependence might be related to the electric Cu$^{1+}$ sublattice in \CCPS since the $B/A$ dependence in Figure~\ref{fig:Adep_Hres}(c) does not reveal any anomaly at the antiferroelectric transition.

\subsection{Summary of properties along a-, b- and c$^*$- axes}
\label{sec:abc_prop}
\begin{table}[!h]
	\centering
	\begin{tabular}{|c|c|c|c|c|c|}
		
	    \hline
		
		  &  & Calculated & Calculated  & Measured & Measured  \\ 
		 
		 axis & $g$-factor & $H\mathrm{_S}$ & $H\mathrm{_{SF}}$ &$H\mathrm{_S}$ &  $H\mathrm{_{SF}}$ \\ \hline
		 
		 ${\bf a}$ &  & 6.5\,T & & $6.7$\,T & \\ \hline
		 ${\bf b}$ & $1.988$ & 6.4\,T & 0.43\,T & $6.6$\,T & $0.43$\,T \\ \hline
		 ${\bf c^*}$ & $1.976$ & 7.8\,T & & $8.5$\,T & \\ \hline

	\end{tabular}
	\label{tab:abc_prop}
\end{table}

\begin{figure*}[!t]
	\centering
	\includegraphics[width=1.0\linewidth]{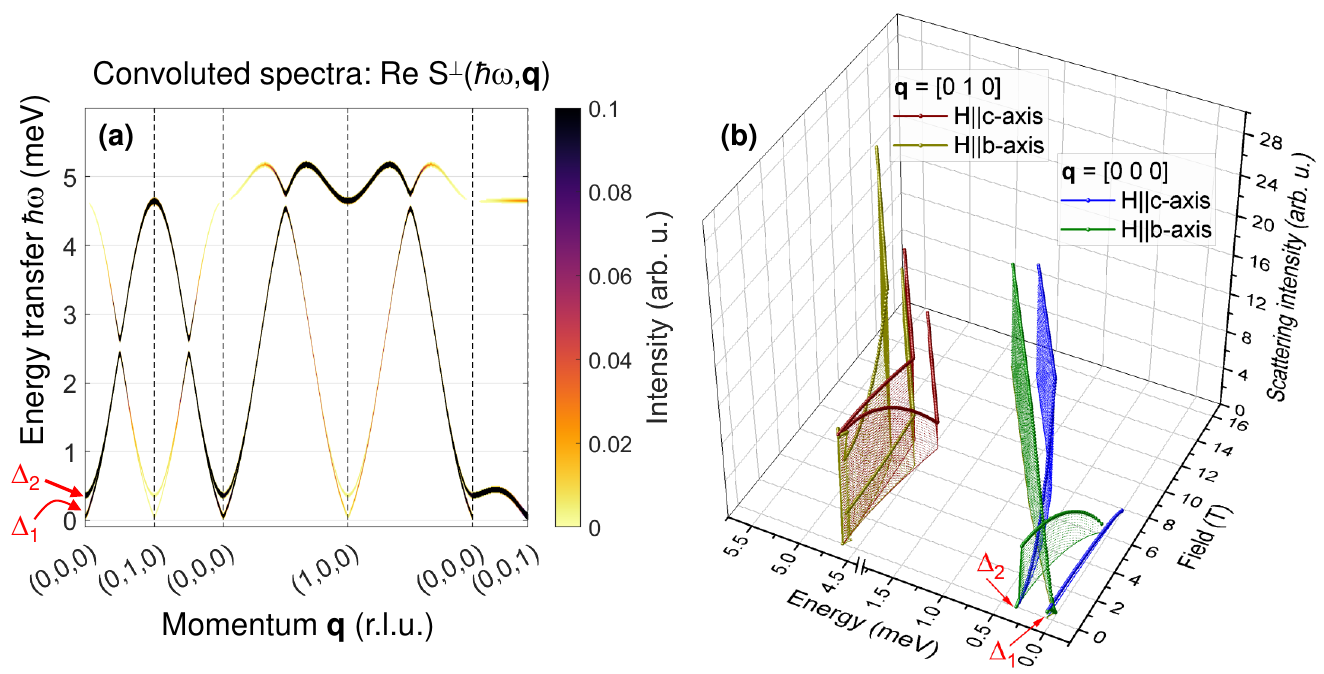}
	\caption{Results of the modeling of the spin wave excitations with the SpinW software \cite{Toth_2015,SpinW}. (a) The spin wave energy as a function of the momentum transfer $\mathbf{q}$. The color code visualizes the inelastic neutron scattering intensity (cross-section) Re\,S$^\perp(\hbar\omega,\mathbf{q}$). (b) The field dependence of the spin wave modes for \HIIb and \HIIc field geometries at $\mathbf{q} =[0,\, 0,\, 0]$ and [0,\,1,\,0]. The zero-field excitation gaps  $\Delta_{1} = 0.054$\,meV (13\,GHz) and  $\Delta_{2}= 0.35$\,meV (84\,GHz) revealed by AFMR are indicated on both panels. }
	\label{fig:SW_dispersion}
\end{figure*}

\subsection{Q-dependence of the spin wave excitations }
\label{sec:spinwaves}

Taking the results of the analysis of the AFMR data in Section~\ref{sec:FDep_3K} and \ref{sec:AFM_ground_state} as input parameters it is possible to model the dispersion of the spin wave excitations in the Brillouin zone of \CCPS with the SpinW software used for the analysis of the inelastic neutron scattering experiments \cite{Toth_2015,SpinW}. The result is shown in Figure~\ref{fig:SW_dispersion}(a). In the zone center $\mathbf{q} = [0,\, 0,\, 0]$ the two modes that are probed by AFMR have distinct energies of 0.054\,meV ($\Delta_{1} = 13$\,GHz) and 0.35\,meV ($\Delta_{2}=84$\,GHz) as indicated in this Figure. Their very steep momentum $\mathbf{q}$-dependent in-plane dispersion with the bandwidth of $\sim 5$\,meV is mostly determined by the strong  intra-plane FM exchange. The anti-crossing gaps of the intersecting spin wave branches seen in Figure~\ref{fig:SW_dispersion}(a) at [0 1/2 0] and [1/2 0 0] arise due to a weak inter-plane AFM coupling $J_{\rm inter}$ and vanish in the limit $J_{\rm inter} = 0$. In contrast, the $\mathbf{q}$-dependence normal to the $ab$-plane, e.g., towards the zone boundary $[0,\, 0,\, 1]$,  is almost flat due to the smallness of $J_{\rm inter}$. This behavior infers  the 2D character of the spin dynamics in \CCPS in agreement with the ESR results (Section~\ref{sec:spin_dynamics}).

The magnetic field dependence of the two modes at zero momentum transfer  $\mathbf{q} = [0,\, 0,\, 0]$ modeled with SpinW closely reproduces the result of the LSWT modeling of the AFMR branches in Section~\ref{sec:FDep_3K} (Figure~\ref{fig:SW_dispersion}(b)). However, the field dependence changes qualitatively at finite momenta as shown in Figure~\ref{fig:SW_dispersion}(b) for the example of $\mathbf{q} = [0,\, 1,\, 0]$ where all modes merge together at $H = 0$ and have the energy of 4.7\,meV. Their degeneracy suggests that at large $\mathbf{q}$ the energies of the spin wave excitations are mainly determined by the isotropic exchange interactions and are much less influenced by anisotropies. This is corroborated by the observation that the field-dependent spin wave branches for \HIIb and \HIIc that cross each other in a field of $\sim 3.5$\,T at $\mathbf{q} = [0,\, 0,\, 0]$ due to a substantial magnetocrystalline anisotropy (Figure~\ref{fig:FDep_3K}(a) and \ref{fig:SW_dispersion}(b)) do not demonstrate this effect at $\mathbf{q} = [0,\, 1,\, 0]$ (Figure~\ref{fig:SW_dispersion}(b)).

\medskip

\bibliography{CuCrP2S6_esr.bib}

@Preamble{
"\providecommand{\noopsort}[1]{}"
# "\providecommand{\singleletter}[1]{#1}%"}

@Article{novoselov2004,
  author    = {Novoselov, K. S. and Geim, A. K. and Morozov, S. V. and Jiang, D. and Zhang, Y. and Dubonos, S. V. and Grigorieva, I. V. and Firsov, A. A.},
  journal   = {Science},
  title     = {Electric field effect in atomically thin carbon films},
  year      = {2004},
  number    = {5696},
  pages     = {666--669},
  volume    = {306},
  doi       = {10.1126/science.1102896},
  publisher = {American Association for the Advancement of Science},
}

@article{chhowalla2013,
	title={The chemistry of two-dimensional layered transition metal dichalcogenide nanosheets},
	author={Chhowalla, M. and Shin, H. S. and Eda, G. and Li, L.-J. and Loh, K. P. and Zhang, H.},
	journal={Nature Chemistry},
	volume={5},
	number={4},
	pages={263--275},
	year={2013},
	publisher={Nature Publishing Group},
	doi={10.1038/nchem.1589}
}

@article{wang2012,
	title={Electronics and optoelectronics of two-dimensional transition metal dichalcogenides},
	author={Wang, Q. H. and Kalantar-Zadeh, K. and Kis, A. and Coleman, J. N. and Strano, M. S.},
	journal={Nature Nanotechnology},
	volume={7},
	number={11},
	pages={699--712},
	year={2012},
	publisher={Nature Publishing Group},
	doi={10.1038/nnano.2012.193}
}

@article{gong2019,
	title={Two-dimensional magnetic crystals and emergent heterostructure devices},
	author={Gong, C. and Zhang, X.},
	journal={Science},
	volume={363},
	number={6428},
	pages={eaav4450},
	year={2019},
	publisher={American Association for the Advancement of Science},
	doi={10.1126/science.aav4450}
}

@article{mak2016,
	title={{Photonics and optoelectronics of 2D semiconductor transition metal dichalcogenides}},
	author={Mak, K. F. and Shan, J.},
	journal={Nature Photonics},
	volume={10},
	number={4},
	pages={216--226},
	year={2016},
	publisher={Nature Publishing Group},
	doi={10.1038/nphoton.2015.282}
}

@article{gong2017,
	title={{Discovery of intrinsic ferromagnetism in two-dimensional van der Waals crystals}},
	author={Gong, C. and Li, L. and Li, Z. and Ji, H. and Stern, A. and Xia, Y. and Cao, T. and Bao, W. and Wang, C. and Wang, Y. and Qiu, Z. Q. and Cava, R. J. and Louie, S. G. and Xia, J. and Zhang, X.},
	journal={Nature},
	volume={546},
	number={7657},
	pages={265--269},
	year={2017},
	publisher={Nature Publishing Group},
	doi={10.1038/nature22060}
}

@article{wang2018,
	title={{Very large tunneling magnetoresistance in layered magnetic semiconductor CrI$_3$}},
	author={Wang, Z. and Gutierrez-Lezama, I. and Ubrig, N. and Kroner, M. and Taniguchi, T. and Watanabe, K. and Imamoglu, A. and Giannini, E. and Morpurgo, A. F.},
	journal={Nature Communications},
	volume={9},
	number={1},
	pages={2516},
	year={2018},
	publisher={Nature Publishing Group},
	doi={10.1038/s41467-018-04953-8}
}

@article{lee2016,
	title={{Ising-type magnetic ordering in atomically thin FePS$_3$}},
	author={Lee, J. and Wang, Z. and Xie, H. and Mak, K. F. and Shan, J.},
	journal={Nano Letters},
	volume={16},
	number={12},
	pages={7433--7438},
	year={2016},
	publisher={American Chemical Society},
	doi={10.1021/acs.nanolett.6b03052}
}

@article{xing2019,
	title = {{Magnon Transport in Quasi-Two-Dimensional van der Waals Antiferromagnets}},
	author = {Xing, Wenyu and Qiu, Luyi and Wang, Xirui and Yao, Yunyan and Ma, Yang and Cai, Ranran and Jia, Shuang and Xie, X. C. and Han, Wei},
	journal = {Phys. Rev. X},
	volume = {9},
	issue = {1},
	pages = {011026},
	numpages = {7},
	year = {2019},
	month = {Feb},
	publisher = {American Physical Society},
	doi = {10.1103/PhysRevX.9.011026},
	url = {https://link.aps.org/doi/10.1103/PhysRevX.9.011026}
}

@ARTICLE{chumak2015,
	title    = "Magnon spintronics",
	author   = "Chumak, A V and Vasyuchka, V I and Serga, A A and Hillebrands, B",
	journal  = "Nature Physics",
	volume   =  11,
	number   =  6,
	pages    = "453--461",
	month    =  jun,
	year     =  2015
}

@ARTICLE{wang2021,
	title    = {{Spin-induced linear polarization of photoluminescence in
	antiferromagnetic van der Waals crystals}},
	author   = "Wang, Xingzhi and Cao, Jun and Lu, Zhengguang and Cohen, Arielle
	and Kitadai, Hikari and Li, Tianshu and Tan, Qishuo and Wilson,
	Matthew and Lui, Chun Hung and Smirnov, Dmitry and Sharifzadeh,
	Sahar and Ling, Xi",
	journal  = "Nature Materials",
	volume   =  20,
	number   =  7,
	pages    = "964--970",
	month    =  jul,
	year     =  2021
}

@ARTICLE{hwangbo2021,
	title    = {{Highly anisotropic excitons and multiple phonon bound states in a
	van der Waals antiferromagnetic insulator}},
	author   = "Hwangbo, Kyle and Zhang, Qi and Jiang, Qianni and Wang, Yong and
	Fonseca, Jordan and Wang, Chong and Diederich, Geoffrey M and
	Gamelin, Daniel R and Xiao, Di and Chu, Jiun-Haw and Yao, Wang
	and Xu, Xiaodong",
	journal  = "Nature Nanotechnology",
	volume   =  16,
	number   =  6,
	pages    = "655--660",
	month    =  jun,
	year     =  2021
}

@article{Selter2023,
	title = {{Crystal growth, exfoliation, and magnetic properties of quaternary quasi-two-dimensional ${\mathrm{CuCrP}}_{2}{\mathrm{S}}_{6}$}},
	author = {Selter, S. and Bestha, K. K. and Bhattacharyya, P. and \"Ozer, B. and Shemerliuk, Y. and Roslova, M. and Vinokurova, E. and Corredor, L. T. and Veyrat, L. and Wolter, A. U. B. and Hozoi, L. and B\"uchner, B. and Aswartham, S.},
	journal = {Phys. Rev. Mater.},
	volume = {7},
	issue = {3},
	pages = {033402},
	numpages = {12},
	year = {2023},
	month = {Mar},
	publisher = {American Physical Society},
	doi = {10.1103/PhysRevMaterials.7.033402},
	}

@Book{Abragam2012,
	title     = {Electron paramagnetic resonance of transition ions},
	publisher = {Oxford University Press, Oxford},
	year      = {2012},
	author    = {Abragam, A. and Bleaney, B.},
}

@Article{wang2023,
	author={Wang, Xiaolei and Shang, Zixuan and Zhang, Chen
	and Kang, Jiaqian and Liu, Tao and Wang, Xueyun and Chen, Siliang and Liu, Haoliang and Tang, Wei and Zeng, Yu-Jia and Guo, Jianfeng and Cheng, Zhihai and Liu, Lei and Pan, Dong and Tong, Shucheng and Wu, Bo and Xie, Yiyang and Wang, Guangcheng and Deng, Jinxiang and Zhai, Tianrui and Deng, Hui-Xiong and Hong, Jiawang and Zhao, Jianhua},
	title={{Electrical and magnetic anisotropies in van der Waals multiferroic CuCrP$_2$S$_6$}},
	journal={Nature Communications},
	year={2023},
	month={Feb},
	day={15},
	volume={14},
	number={1},
	pages={840},
	issn={2041-1723},
	doi={10.1038/s41467-023-36512-1},
	url={https://doi.org/10.1038/s41467-023-36512-1}
}

@Book{Turov,
  author    = {E. A. Turov},
  publisher = {Academic Press, New York},
  title     = {Physical Properties of Magnetically\allowbreak\ Ordered\allowbreak\ Crystals},
  year      = {1965},
}

@article{Holstein1940,
	title = {Field Dependence of the Intrinsic Domain Magnetization of a Ferromagnet},
	author = {Holstein, T. and Primakoff, H.},
	journal = {Phys. Rev.},
	volume = {58},
	issue = {12},
	pages = {1098--1113},
	numpages = {0},
	year = {1940},
	month = {Dec},
	publisher = {American Physical Society},
	doi = {10.1103/PhysRev.58.1098},
	url = {https://link.aps.org/doi/10.1103/PhysRev.58.1098}
}

@article{alfonsov2021,
	title = {{Magnetic-field tuning of the spin dynamics in the magnetic topological insulators (MnBi$_{2}$Te$_{4}$)(Bi$_{2}$Te$_{3}$)$_{n}$}},
	author = {Alfonsov, A. and Mehlawat, K. and Zeugner, A. and Isaeva, A. and B\"uchner, B. and Kataev, V.},
	journal = {Phys. Rev. B},
	volume = {104},
	issue = {19},
	pages = {195139},
	numpages = {12},
	year = {2021},
	month = {Nov},
	publisher = {American Physical Society},
	doi = {10.1103/PhysRevB.104.195139},
	url = {https://link.aps.org/doi/10.1103/PhysRevB.104.195139}
}

@article{abraham2023,
	title = {{Magnetic anisotropy and low-energy spin dynamics in the van der Waals compounds Mn$_2$P$_2$S$_6$ and MnNiP$_2$S$_6$}},
	author = {Abraham, J. J. and Senyk, Y. and Shemerliuk, Y. and Selter, S. and Aswartham, S. and B\"uchner, B. and Kataev, V. and Alfonsov, A.},
	journal = {Phys. Rev. B},
	volume = {107},
	issue = {16},
	pages = {165141},
	numpages = {11},
	year = {2023},
	month = {Apr},
	publisher = {American Physical Society},
	doi = {10.1103/PhysRevB.107.165141},
	url = {https://link.aps.org/doi/10.1103/PhysRevB.107.165141}
}

@article{senyk2023,
	title = {{Evolution of the spin dynamics in the van der Waals system $M_2$P$_2$S$_6$ ($M_2$=Mn$_2$, MnNi, Ni$_2$) series probed by electron spin resonance spectroscopy}},
	author = {Senyk, Y. and Abraham, J. J. and Shemerliuk, Y. and Selter, S. and Aswartham, S. and B\"uchner, B. and Kataev, V. and Alfonsov, A.},
	journal = {Phys. Rev. Mater.},
	volume = {7},
	issue = {1},
	pages = {014003},
	numpages = {9},
	year = {2023},
	month = {Jan},
	publisher = {American Physical Society},
	doi = {10.1103/PhysRevMaterials.7.014003},
	url = {https://link.aps.org/doi/10.1103/PhysRevMaterials.7.014003}
}

@article{mehlawat2022,
	title = {{Low-energy excitations and magnetic anisotropy of the layered van der Waals antiferromagnet Ni$_2$P$_2$S$_6$}},
	author = {Mehlawat, K. and Alfonsov, A. and Selter, S. and Shemerliuk, Y. and Aswartham, S. and B\"uchner, B. and Kataev, V.},
	journal = {Phys. Rev. B},
	volume = {105},
	issue = {21},
	pages = {214427},
	numpages = {7},
	year = {2022},
	month = {Jun},
	publisher = {American Physical Society},
	doi = {10.1103/PhysRevB.105.214427},
	url = {https://link.aps.org/doi/10.1103/PhysRevB.105.214427}
}

@article{zeisner2019,
	title = {{Magnetic anisotropy and spin-polarized two-dimensional electron gas in the van der Waals ferromagnet Cr$_2$Ge$_2$Te$_6$}},
	author = {Zeisner, J. and Alfonsov, A. and Selter, S. and Aswartham, S. and Ghimire, M. P. and Richter, M. and van den Brink, J. and B\"uchner, B. and Kataev, V.},
	journal = {Phys. Rev. B},
	volume = {99},
	issue = {16},
	pages = {165109},
	numpages = {14},
	year = {2019},
	month = {Apr},
	publisher = {American Physical Society},
	doi = {10.1103/PhysRevB.99.165109},
	url = {https://link.aps.org/doi/10.1103/PhysRevB.99.165109}
}

@article{maisonneuve1995,
	title = {{On CuCrP$_2$S$_6$: Copper disorder, stacking distortions, and magnetic ordering}},
	journal = {Journal of Solid State Chemistry},
	volume = {116},
	number = {1},
	pages = {208-210},
	year = {1995},
	issn = {0022-4596},
	doi = {https://doi.org/10.1006/jssc.1995.1204},
	url = {https://www.sciencedirect.com/science/article/pii/S0022459685712042},
	author = {V. Maisonneuve and C. Payen and V.B. Cajipe},
}

@Article{park2022,
  author   = {Park, Chang Bae and Shahee, Aga and Kim, Kwang-Tak and Patil, Deepak R. and Guda, Sergey Alexandrovich and Ter-Oganessian, Nikita and Kim, Kee Hoon},
  journal  = {Advanced Electronic Materials},
  title    = {{Observation of spin-induced ferroelectricity in a layered van der Waals antiferromagnet CuCrP$_2$S$_6$}},
  year     = {2022},
  number   = {6},
  pages    = {2101072},
  volume   = {8},
  doi      = {https://doi.org/10.1002/aelm.202101072},
  url      = {https://onlinelibrary.wiley.com/doi/abs/10.1002/aelm.202101072},
}

@Article{lai2019,
	author ="Lai, Youfang and Song, Zhigang and Wan, Yi and Xue, Mingzhu and Wang, Changsheng and Ye, Yu and Dai, Lun and Zhang, Zhidong and Yang, Wenyun and Du, Honglin and Yang, Jinbo",
	title  ={{Two-dimensional ferromagnetism and driven ferroelectricity in van der Waals CuCrP$_2$S$_6$}},
	journal  ="Nanoscale",
	year  ="2019",
	volume  ="11",
	issue  ="12",
	pages  ="5163-5170",
	publisher  ="The Royal Society of Chemistry",
	doi  ="10.1039/C9NR00738E",
	url  ="http://dx.doi.org/10.1039/C9NR00738E",
}

@Article{pal2024,
  author   = {Pal, Riju and Abraham, Joyal John and Mistonov, Alexander and Mishra, Swarnamayee and Stilkerich, Nina and Mondal, Suchanda and Mandal, Prabhat and Pal, Atindra Nath and Geck, Jochen and Büchner, Bernd and Kataev, Vladislav and Alfonsov, Alexey},
  journal  = {Advanced Functional Materials},
  title    = {Disentangling the Unusual Magnetic Anisotropy of the Near-Room-Temperature Ferromagnet {Fe$_4$GeTe$_2$}},
  year     = {2024},
  number   = {38},
  pages    = {2402551},
  volume   = {34},
  doi      = {https://doi.org/10.1002/adfm.202402551},
  url      = {https://onlinelibrary.wiley.com/doi/abs/10.1002/adfm.202402551},
}

@Article{maisonneuve1993,
	author={Maisonneuve, V.
	and Cajipe, V. B.
	and Payen, C.},
	title={Low-temperature neutron powder diffraction study of copper chromium thiophosphate {CuCrP$_2$S$_6$}: observation of an ordered, antipolar copper sublattice},
	journal={Chemistry of Materials},
	year={1993},
	month={Jun},
	day={01},
	publisher={American Chemical Society},
	volume={5},
	number={6},
	pages={758-760},
	issn={0897-4756},
	doi={10.1021/cm00030a006},
	url={https://doi.org/10.1021/cm00030a006}
}

@Article{io2023,
  author  = {Io, Weng Fu and Pang, Sin -Yi and Wong, Lok Wing and Zhao, Yuqian and Ding, Ran and Mao, Jianfeng and Zhao, Yifei and Guo, Feng and Yuan, Shuoguo and Zhao, Jiong and Yi, Jiabao and Hao, Jianhua},
  journal = {Nature Communications},
  title   = {{Direct observation of intrinsic room-temperature ferroelectricity in 2D layered CuCrP$_2$S$_6$}},
  year    = {2023},
  issn    = {2041-1723},
  month   = {Nov},
  number  = {1},
  pages   = {7304},
  volume  = {14},
  day     = {11},
  doi     = {10.1038/s41467-023-43097-2},
  url     = {https://doi.org/10.1038/s41467-023-43097-2},
}

@article{Colombet1982,
	title = {{Structural aspects and magnetic properties of the lamellar compound Cu$_{0.50}$Cr$_{0.50}$PS$_3$}},
	journal = {J. Solid State Chem.},
	volume = {41},
	number = {2},
	pages = {174-184},
	year = {1982},
	issn = {0022-4596},
	doi = {https://doi.org/10.1016/0022-4596(82)90200-6},
	author = {P. Colombet and A. Leblanc and M. Danot and J. Rouxel},
	}

@Article{Susner2020,
  author   = {Susner, M. A. and Rao, R. and Pelton, A. T. and McLeod, M. V. and Maruyama, B.},
  journal  = {Phys. Rev. Mater.},
  title    = {{Temperature-dependent Raman scattering and x-ray diffraction study of phase transitions in layered multiferroic $\mathrm{CuCr}{\mathrm{P}}_{2}{\mathrm{S}}_{6}$}},
  year     = {2020},
  month    = {Oct},
  pages    = {104003},
  volume   = {4},
  doi      = {10.1103/PhysRevMaterials.4.104003},
  issue    = {10},
  numpages = {10},
}

@article{Cho2022,
	Author = {Cho, Kwanghee and Lee, Seungyeol and Kalaivanan, Raju and Sankar, Raman
	and Choi, Kwang-Yong and Park, Soonyong},
	Title = {{Tunable Ferroelectricity in Van der Waals Layered Antiferroelectric
	CuCrP$_2$S$_6$}},
	Journal = {Adv. Func. Mater.},
	Year = {2022},
	Volume = {32},
	Number = {36},
	Month = {SEP},
	DOI = {10.1002/adfm.202204214},
}

@article{Kleemann2011,
	title = {{Magnetic and polar phases and dynamical clustering in multiferroic layered solid solutions CuCr${}_{1\ensuremath{-}x}$In${}_{x}$P${}_{2}$S${}_{6}$}},
	author = {Kleemann, W. and Shvartsman, V. V. and Borisov, P. and Banys, J. and Vysochanskii, Yu. M.},
	journal = {Phys. Rev. B},
	volume = {84},
	issue = {9},
	pages = {094411},
	numpages = {8},
	year = {2011},
	month = {Sep},
	doi = {10.1103/PhysRevB.84.094411},
}

@article{Tang2023,
	Author = {Tang, Chunli and Alahmed, Laith and Mahdi, Muntasir and Xiong, Yuzan and
	Inman, Jerad and McLaughlin, Nathan J. and Zollitsch, Christoph and Kim,
	Tae Hee and Du, Chunhui Rita and Kurebayashi, Hidekazu and Santos, Elton
	J. G. and Zhang, Wei and Li, Peng and Jin, Wencan},
	Title = {{Spin dynamics in van der Waals magnetic systems}},
	Journal = {Phys. Rep.},
	Year = {2023},
	Volume = {1032},
	Pages = {1-36},
	Month = {AUG 24},
	DOI = {10.1016/j.physrep.2023.09.002},
}

@article{Kataev2024,
	Author = {Kataev, Vladislav and Buechner, Bernd and Alfonsov, Alexey},
	Title = {{Electron Spin Resonance Spectroscopy on Magnetic van der Waals Compounds}},
	Journal = {Appl. Magn. Reson.},
	Year = {2024},
	Volume = {55},
	Number = {9, SI},
	Pages = {923-960},
	Month = {SEP},
	DOI = {10.1007/s00723-024-01671-x},
	EarlyAccessDate = {JUL 2024},
	ISSN = {0937-9347},
	EISSN = {1613-7507},
	ResearcherID-Numbers = {Kataev, Vladislav/HLQ-6390-2023},
	Unique-ID = {WOS:001260430100001},
}

@article{Kobets2009,
	author = {Kobets, M. I. and Dergachev, K. G. and Gnatchenko, S. L. and Khats’ko, E. N. and Vysochanskii, Yu. M. and Gurzan, M. I.},
	title = {{Antiferromagnetic resonance in Mn$_2$P$_2$S$_6$}},
	journal = {Low Temp. Phys.},
	volume = {35},
	number = {12},
	pages = {930-934},
	year = {2009},
	month = {12},
	doi = {10.1063/1.3272560},
}

@article{Wyzula2022,
	Author = {Wyzula, Jan and Mohelsky, Ivan and Vaclaykoya, Diana and Kapuscinski,
	Piotr and Veis, Martin and Faugeras, Clement and Potemski, Marek and
	Zhitomirsky, Mike E. and Orlita, Milan},
	Title = {{High-Angular Momentum Excitations in Collinear Antiferromagnet FePS$_3$}},
	Journal = {Nano Lett.},
	Year = {2022},
	Volume = {22},
	Number = {23},
	Pages = {9741-9747},
	Month = {DEC 14},
	DOI = {10.1021/acs.nanolett.2c04111},
}

@article{Flebus2023,
	author = {Flebus, B. and Rezende, S. M. and Grundler, D. and Barman, A.},
	title = "{Recent advances in magnonics}",
	journal = {J. Appl. Phys.},
	volume = {133},
	number = {16},
	pages = {160401},
	year = {2023},
	month = {04},
	doi = {10.1063/5.0153424},
}

@InCollection{Benner1990,
  author    = {Benner, H. and Boucher, J.P.},
  booktitle = {Magnetic Properties of Layered Transition Metal Compounds},
  publisher = {Kluwer Academic Publishers, Dordrecht},
  title     = {{Spin Dynamics in the Paramagnetic Regime: NMR and EPR in Two-Dimensional Magnets}},
  year      = {1990},
  chapter   = {Spin Dynamics in the Paramagnetic Regime: NMR and EPR in Two-Dimensional Magnets},
  editor    = {de Jongh, L.J.},
  pages     = {323-378},
  doi       = {10.1007/978-94-009-1860-3},
}

@article{Richards1974,
	title = {Exchange narrowing of electron spin resonance in a two-dimensional system},
	author = {Richards, Peter M. and Salamon, M. B.},
	journal = {Phys. Rev. B},
	volume = {9},
	issue = {1},
	pages = {32--45},
	numpages = {0},
	year = {1974},
	month = {Jan},
	publisher = {American Physical Society},
	doi = {10.1103/PhysRevB.9.32},
	url = {https://link.aps.org/doi/10.1103/PhysRevB.9.32}
}

@article{Benner1978,
	title = {{Experimental evidence for spin diffusion in the quasi-two-dimensional Heisenberg paramagnet ${({\mathrm{C}}_{2}{\mathrm{H}}_{5}\mathrm{N}{\mathrm{H}}_{3})}_{2}$Mn${\mathrm{Cl}}_{4}$}},
author = {Benner, H.},
journal = {Phys. Rev. B},
volume = {18},
issue = {1},
pages = {319--325},
numpages = {0},
year = {1978},
month = {Jul},
publisher = {American Physical Society},
doi = {10.1103/PhysRevB.18.319},
url = {https://link.aps.org/doi/10.1103/PhysRevB.18.319}
}

@Misc{peak_splitting,
  note = {{In certain microwave frequency ranges, some instrumental effects distorting the lineshape of the ESR signal in the form of multiple closely lying resonances may occur in the used HF-ESR setup. They are due to the closeness of the microwave wavelength to the size of the sample flake being measured. At temperatures where the sample's magnetization is strong, this kind of splitting of the signal becomes pronounced, as is the case of CuCrP$_2$S$_6$ at temperatures $T \lesssim 60$\,K.}},
}

@article{Gomonay2018,
	Author = {Gomonay, O. and Baltz, V. and Brataas, A. and Tserkovnyak, Y.},
	Title = {Antiferromagnetic spin textures and dynamics},
	Journal = {Nat. Phys.},
	Year = {2018},
	Volume = {14},
	Number = {3},
	Pages = {213-216},
	Month = {MAR},
	DOI = {10.1038/s41567-018-0049-4},
	url = {https://doi.org/10.1038/s41567-018-0049-4}
	}

@article{Han2023,
	Author = {Han, Jiahao and Cheng, Ran and Liu, Luqiao and Ohno, Hideo and Fukami,
	Shunsuke},
	Title = {Coherent antiferromagnetic spintronics},
	Journal = {Nat. Mater.},
	Year = {2023},
	Volume = {22},
	Number = {6},
	Pages = {684-695},
	Month = {JUN},
	DOI = {10.1038/s41563-023-01492-6},
	url = {https://www.nature.com/articles/s41563-023-01492-6}
	}

@Article{Rezende2016,
  author    = {Rezende, S. M. and Rodr\'{\i}guez-Su\'arez, R. L. and Azevedo, A.},
  journal   = {Phys. Rev. B},
  title     = {Diffusive magnonic spin transport in antiferromagnetic insulators},
  year      = {2016},
  month     = {Feb},
  pages     = {054412},
  volume    = {93},
  doi       = {10.1103/PhysRevB.93.054412},
  issue     = {5},
  numpages  = {11},
  publisher = {American Physical Society},
  url       = {https://link.aps.org/doi/10.1103/PhysRevB.93.054412},
}

@Article{Rezende2019,
  author   = {Rezende, Sergio M. and Azevedo, Antonio and Rodríguez-Suárez, Roberto L.},
  journal  = {Journal of Applied Physics},
  title    = {Introduction to antiferromagnetic magnons},
  year     = {2019},
  issn     = {0021-8979},
  month    = {10},
  number   = {15},
  pages    = {151101},
  volume   = {126},
  doi      = {10.1063/1.5109132},
  url      = {https://doi.org/10.1063/1.5109132},
}

@Article{Danisch2021,
  author    = {Simon Danisch and Julius Krumbiegel},
  journal   = {Journal of Open Source Software},
  title     = {{Makie.jl}: Flexible high-performance data visualization for {Julia}},
  year      = {2021},
  number    = {65},
  pages     = {3349},
  volume    = {6},
  doi       = {10.21105/joss.03349},
  publisher = {The Open Journal},
  url       = {https://doi.org/10.21105/joss.03349},
}

@Article{Bezanson2017,
  author   = {Bezanson, Jeff and Edelman, Alan and Karpinski, Stefan and Shah, Viral B.},
  journal  = {SIAM Review},
  title    = {Julia: A Fresh Approach to Numerical Computing},
  year     = {2017},
  number   = {1},
  pages    = {65-98},
  volume   = {59},
  doi      = {10.1137/141000671},
  eprint   = {https://doi.org/10.1137/141000671},
  url      = {https://doi.org/10.1137/141000671},
}

@misc{SpinW, 
	doi = {10.5281/zenodo.2651100}, 
	url = {https://zenodo.org/record/2651100}, 
	author = {Tóth, Sándor and Ward, Simon}, 
	title = {{SpinW/spinw: SpinW 3.1, Zenodo}}, 
	publisher = {Zenodo}, year = {2019} }

@article{Toth_2015,
	doi = {10.1088/0953-8984/27/16/166002},
	url = {https://dx.doi.org/10.1088/0953-8984/27/16/166002},
	year = {2015},
	month = {mar},
	publisher = {IOP Publishing},
	volume = {27},
	number = {16},
	pages = {166002},
	author = {S Toth and B Lake},
	title = {{Linear spin wave theory for single-Q incommensurate magnetic structures}},
	journal = {J. Phys.: Condens. Matter}
}

@book{APEX4,
	title = {{APEX4 Crystallography Software Suite}},
	publisher = {Bruker AXS Inc.: Madison, WI, USA},
	year = {2021}
}

@Misc{axes_avsb,
  note = {{In the literature there was no certainty in the definition of the magnetically easy axis. Theoretical calculation in \cite{lai2019} suggest the $a$-axis to be the easy axis. However, in \cite{wang2023}, where the easy axis is identified experimentally, two different figures show opposing results. In one the $a$-axis is the easy axis and in the other figure it is the $b$-axis. Therefore, we have conducted X-ray single crystal diffraction measurements in order to determine the correspondence between crystallographic and magnetic axes. The sample with identified axes was measured in the ESR spectrometer and in the SQUID magnetometer. As a result, the spin flop corresponding to the easy direction was observed when the magnetic field was parallel to the $b$-axis.}},
}

@Article{Koepernik1999full,
  author    = {Koepernik, Klaus and Eschrig, Helmut},
  journal   = {Phys. Rev. B},
  title     = {{Full-potential nonorthogonal local-orbital minimum-basis band-structure scheme}},
  year      = {1999},
  number    = {3},
  pages     = {1743},
  volume    = {59},
  publisher = {APS},
}

@Article{PBE96,
  author  = {Perdew, John P. and Burke, Kieron and Ernzerhof, Matthias},
  title   = {Generalized Gradient Approximation Made Simple},
  journal = {Phys. Rev. Lett.},
  year    = {1996},
  volume  = {77},
  pages   = {3865--3868},
  doi     = {10.1103/PhysRevLett.77.3865},
}

@article{SCAN,
  title = {{Strongly Constrained and Appropriately Normed Semilocal Density Functional}},
  author = {Sun, Jianwei and Ruzsinszky, Adrienn and Perdew, John P.},
  journal = {Phys. Rev. Lett.},
  volume = {115},
  pages = {036402},
  year = {2015},
  doi = {10.1103/PhysRevLett.115.036402},
}

@Article{VASP,
  author  = {Kresse, G. and Furthm{\"{u}}ller, J.},
  title   = {Efficient iterative schemes for {\textit{ab initio}} total-energy calculations using a plane-wave basis set},
  journal = {Phys. Rev. B},
  year    = {1996},
  volume  = {54},
  pages   = {11169--11186},
  doi     = {10.1103/PhysRevB.54.11169},
}

@Article{VASP_2,
  author  = {Kresse, G. and Furthm{\"{u}}ller, J.},
  title   = {Efficiency of ab-initio total energy calculations for metals and semiconductors using a plane-wave basis set},
  journal = {Comput. Mater. Sci.},
  year    = {1996},
  volume  = {6},
  pages   = {15--50},
  doi     = {10.1016/0927-0256(96)00008-0},
}

@article{VASP_PAW,
  title = {From ultrasoft pseudopotentials to the projector augmented-wave method},
  author = {Kresse, G. and Joubert, D.},
  journal = {Phys. Rev. B},
  volume = {59},
  issue = {3},
  pages = {1758--1775},
  year = {1999},
  doi = {10.1103/PhysRevB.59.1758},
}

@Article{Vesta,
  author   = {Momma, Koichi and Izumi, Fujio},
  journal  = {Journal of Applied Crystallography},
  title    = {{{\it VESTA3} for three-dimensional visualization of crystal, volumetric and morphology data}},
  year     = {2011},
  month    = {Dec},
  number   = {6},
  pages    = {1272--1276},
  volume   = {44},
  doi      = {10.1107/S0021889811038970},
  url      = {https://doi.org/10.1107/S0021889811038970},
}

@Article{Moro2022,
  author   = {Moro, F. and Ke, S. and del Águila, A. G. and Söll, A. and Sofer, Z. and Wu, Q. and Yue, M. and Li, L. and Liu, X. and Fanciulli, M.},
  journal  = {Advanced Functional Materials},
  title    = {{Revealing 2D Magnetism in a Bulk {CrSBr} Single Crystal by Electron Spin Resonance}},
  year     = {2022},
  number   = {45},
  pages    = {2207044},
  volume   = {32},
  doi      = {https://doi.org/10.1002/adfm.202207044},
  url      = {https://advanced.onlinelibrary.wiley.com/doi/abs/10.1002/adfm.202207044},
}

@Article{Kim2019,
  author   = {Kim, Kangwon and Lim, Soo Yeon and Lee, Jae-Ung and Lee, Sungmin and Kim, Tae Yun and Park, Kisoo and Jeon, Gun Sang and Park, Cheol-Hwan and Park, Je-Geun and Cheong, Hyeonsik},
  journal  = {Nature Communications},
  title    = {{Suppression of magnetic ordering in XXZ-type antiferromagnetic monolayer NiPS$_3$}},
  year     = {2019},
  issn     = {2041-1723},
  month    = {Jan},
  number   = {1},
  pages    = {345},
  volume   = {10},
  day      = {21},
  doi      = {10.1038/s41467-018-08284-6},
  url      = {https://doi.org/10.1038/s41467-018-08284-6},
}

@article{Manas2025,
	title = {{Fundamentals and applications of van der Waals magnets in magnon spintronics}},
	journal = {Newton},
	volume = {1},
	number = {1},
	pages = {100018},
	year = {2025},
	doi = {https://doi.org/10.1016/j.newton.2025.100018},
	url = {https://www.sciencedirect.com/science/article/pii/S2950636025000106},
	author = {Samuel Ma\~{n}as-Valero and Toeno {van der Sar} and Rembert A. Duine and Bart {van Wees}}
	}

@article{Han2020,
	Author = {Han, Jiahao and Zhang, Pengxiang and Bi, Zhen and Fan, Yabin and Safi,
	Taqiyyah S. and Xiang, Junxiang and Finley, Joseph and Fu, Liang and
	Cheng, Ran and Liu, Luqiao},
	Title = {{Birefringence-like spin transport via linearly polarized
	antiferromagnetic magnons}},
	Journal = {Nat. Nanotechnol.},
	Year = {2020},
	Volume = {15},
	Number = {7},
	Pages = {563},
	Month = {JUL},
	DOI = {10.1038/s41565-020-0703-8}	
}

@article{Yang2021,
	Author = {Yang, Shengxue and Zhang, Tianle and Jiang, Chengbao},
	Title = {{van der Waals Magnets: Material Family, Detection and Modulation of
	Magnetism, and Perspective in Spintronics}},
	Journal = {Adv. Sci.},
	Year = {2021},
	Volume = {8},
	Number = {2},
	Month = {JAN},
	DOI = {10.1002/advs.202002488},
}

@article{Khan2020,
	title = {{Recent breakthroughs in two-dimensional van der Waals magnetic materials and emerging applications}},
	journal = {Nano Today},
	volume = {34},
	pages = {100902},
	year = {2020},
	doi = {https://doi.org/10.1016/j.nantod.2020.100902},
	url = {https://www.sciencedirect.com/science/article/pii/S1748013220300712},
	author = {Yahya Khan and Sk. Md. Obaidulla and Mohammad Rezwan Habib and Anabil Gayen and Tao Liang and Xuefeng Wang and Mingsheng Xu},
	}

@article{Burch2018,
	Author = {Burch, Kenneth S. and Mandrus, David and Park, Je-Geun},
	Title = {{Magnetism in two-dimensional van der Waals materials}},
	Journal = {Nature},
	Year = {2018},
	Volume = {563},
	Number = {7729},
	Pages = {47-52},
	Month = {NOV 1},
	DOI = {10.1038/s41586-018-0631-z},	
}

@article{Okuda1986,
	author = {Okuda ,Kiichi and Kurosawa ,Ko and Saito ,Shozo and Honda ,Makoto and Yu ,Zhihong and Date ,Muneyuki},
	title = {{Magnetic Properties of Layered Compound MnPS$_3$}},
	journal = {J. Phys. Soc. of Jpn},
	volume = {55},
	number = {12},
	pages = {4456-4463},
	year = {1986},
	doi = {10.1143/JPSJ.55.4456},
}

@Article{Rogic2024,
  author   = {{L. Rogi\'c and N. Somun and S. Griffitt and A. Najev and M. Spai\'c and S. Hameed and Y. Shemerliuk and S. Aswartham and M. Orlita and A. Alfonsov and D. Pelc}},
  journal  = {Review of Scientific Instruments},
  title    = {{Cryogenic continuous-wave optical spectrometer for sub-THz frequencies}},
  year     = {2025},
  issn     = {0034-6748},
  month    = {08},
  number   = {8},
  pages    = {083101},
  volume   = {96},
  doi      = {10.1063/5.0251272},
  url      = {https://doi.org/10.1063/5.0251272},
}

\end{document}